\documentclass{aastex63}
\usepackage{lineno}
\usepackage{graphicx}
\usepackage{subfigure}
\usepackage{color}
\usepackage{amsmath}
\submitjournal{ApJ}
\shorttitle{Late afterglow bump/plateau around the jet break}
\shortauthors{Li et al.}
\newcommand{\MyFigA}{\ref{MyFigA}}
\newcommand{\MyFigB}{\ref{MyFigB}}
\newcommand{\MyFigC}{\ref{MyFigC}}
\newcommand{\MyFigD}{\ref{MyFigD}}
\newcommand{\MyFigE}{\ref{MyFigE}}
\newcommand{\MyFigF}{\ref{MyFigF}}
\newcommand{\MyFigG}{\ref{MyFigG}}
\newcommand{\MyFigH}{\ref{MyFigH}}
\newcommand{\MyTabA}{\ref{MyTabA}}
\newcommand{\MyTabB}{\ref{MyTabB}}
\begin{document}
\title{Late Afterglow Bump/Plateau around the Jet Break: Signature of a free-to-shocked wind Environment in Gamma-ray Burst}
\correspondingauthor{Da-Bin Lin}
\email{lindabin@gxu.edu.cn}
\author{Xiao-Yan Li}
\affil{Laboratory for Relativistic Astrophysics, Department of Physics, Guangxi University, Nanning 530004, China}
\author{Da-Bin Lin}
\affil{Laboratory for Relativistic Astrophysics, Department of Physics, Guangxi University, Nanning 530004, China}
\author{Jia Ren}
\affil{Laboratory for Relativistic Astrophysics, Department of Physics, Guangxi University, Nanning 530004, China}
\author{Shu-Jin Hou}
\affil{College of Physics and Electronic Engineering, Nanyang Normal University, Nanyang, Henan 473061, China}
\affil{Laboratory for Relativistic Astrophysics, Department of Physics, Guangxi University, Nanning 530004, China}
\author{Yu-Fei Li}
\affil{Laboratory for Relativistic Astrophysics, Department of Physics, Guangxi University, Nanning 530004, China}
\author{Xiang-Gao Wang}
\affil{Laboratory for Relativistic Astrophysics, Department of Physics, Guangxi University, Nanning 530004, China}
\author{En-Wei Liang}
\affil{Laboratory for Relativistic Astrophysics, Department of Physics, Guangxi University, Nanning 530004, China}
\begin{abstract}
A number of gamma-ray bursts (GRBs)
exhibit the late simultaneous bumps in their optical and X-ray afterglows around the jet break. Its origin is unclear.
Based on the following two facts,
we suggest that this feature may sound a transition of circum-burst environment
from a free-wind medium to a homogeneous medium.
(I) The late bump followed by a steep decay is strongly reminiscent of the afterglows of GRB~170817A,
which is attributed to an off-axis observed external-forward shock (eFS) propagating in an interstellar medium.
(II) Observations seem to feature a long shallow decay before the late optical bump,
which is different from the afterglow of GRB~170817A.
In this paper, we study the emission of an eFS propagating in a  free-to-shocked wind
for on/off-axis observers, where the mass density in the shocked-wind is almost constant.
The late simultaneous bumps/plateaux in the optical and X-ray afterglows are really found
around the jet break for high-viewing-angle observers.
Moreover, there is a long plateau or shallow decay before the late bump in the theoretical light-curves,
which is formed during the eFS propagating in the free-wind.
For low-viewing-angle observers,
the above bumps appear only in the situation that
the structured jet has a low characteristic angle
and the deceleration radius of the on-axis jet flow is at around or beyond the free-wind boundary.
As examples, the X-ray and optical afterglows of GRBs~120326A, 120404A, and 100814A are fitted.
We find that an off-axis observed eFS in a free-to-shocked wind
can well explain the afterglows in these bursts.
\end{abstract}

\section{Introduction}\label{Sec:Intro}
Gamma-ray bursts (GRBs) are the most luminous sources of electromagnetic radiation known in the Universe.
Observationally, GRBs generally appear as a brief and intense $\gamma$-rays
followed by a long-lived afterglow emission.
With the observations of \emph{Swift} satellite,  the diversity of light-curves in X-ray afterglows have been revealed
(\citealp{Gehrels_N-2004-Chincarini_G-ApJ.611.1005G}; \citealp{Burrows_DN-2005-Hill_JE-SSRv.120.165B}).
However, a canonical light-curve consisting of four power-law segments and a flaring component
has been suggested, i.e.,
an  initial steep decay with a typical slope $\lesssim -3$,
a shallow decay with a typical slope $\sim -0.5$,
a normal decay with a typical slope $\sim -1.2$,
a late steep decay with a typical slope $\sim -2$,
and
one or several X-ray flares
(\citealp{Zhang_B-2006-Fan_YZ-ApJ.642.354Z}; \citealp{Nousek_JA-2006-Kouveliotou_C-ApJ.642.389N}; \citealp{OBrien_PT-2006-Willingale_R-ApJ.647.1213O}; \citealp{Zhang_BB-2007-Liang_EW-ApJ.666.1002Z}).
The initial steep decay component is believed to be the tail
of the prompt emission due to the ``curvature effect'' of the high-latitude emission
(\citealp{Barthelmy_SD-2005-Cannizzo_JK-ApJ.635L.133B};
\citealp{Liang_EW-2006-Zhang_B-ApJ.646.351L};
\citealp{OBrien_PT-2006-Willingale_R-ApJ.647.1213O};
\citealp{Lin_DB-2017-Mu_HJ-ApJ.840.118L,Lin_DB-2017-Mu_HJ-ApJ.840.95L}).
The shallow, the normal, and the late steep decays are always attributed to the external shock
(\citealp{Zhang_B-2006-Fan_YZ-ApJ.642.354Z}; \citealp{Nousek_JA-2006-Kouveliotou_C-ApJ.642.389N}; \citealp{Panaitescu_A-2006-Meszaros_P-MNRAS.369.2059P}),
which is developed when a relativistic jet propagates into the circum-burst medium.
If there is no energy injection into the external shock,
a normal decay phase would appear in the afterglows.
However, the decay may become shallow during the phase with continuous energy injection
into the external shock.
The late steep decay corresponds to the emission of the external shock after the jet break.
The X-ray flares generally show a sharp rise with a steep decay
and thus could not be produced in the external shock (e.g., \citealp{Romano_P-2006-Moretti_A-A&A.450.59R}; \citealp{Falcone_AD-2006-Burrows_DN-ApJ.641.1010F}).
It is suggested that most of X-ray flares have the same physical origin as the prompt $\gamma$-rays
(\citealp{Burrows_DN-2005-Romano_P-Sci.309.1833B}; \citealp{Falcone_AD-2006-Burrows_DN-ApJ.641.1010F,Falcone_AD-2007-Morris_D-ApJ.671.1921F};
\citealp{Zhang_B-2006-Fan_YZ-ApJ.642.354Z};
\citealp{Nousek_JA-2006-Kouveliotou_C-ApJ.642.389N}; \citealp{Liang_EW-2006-Zhang_B-ApJ.646.351L};
\citealp{Chincarini_G-2007-Moretti_A-ApJ.671.1903C,Chincarini_G-2010-Mao_J-MNRAS.406.2113C};
\citealp{Hou_SJ-2014-Geng_JJ-ApJ.785.113H};
\citealp{Wu_XF-2013-Hou_SJ-ApJ.767L.36W};
\citealp{Yi_SX-2015-Wu_XF-ApJ.807.92Y,Yi_SX-2016-Xi_SQ-ApJS.224.20Y}; \citealp{Mu_HJ-2016-Gu_WM-ApJ.832.161M,Mu_HJ-2016-Lin_DB-ApJ.831.111M}).

Multi-wavelength observations of afterglows have also revealed some puzzling features,
which are beyond the expectation from the simple standard external shock scenario
(\citealp{Panaitescu_A-2006-Meszaros_P-MNRAS.369.2059P};
\citealp{Panaitescu_A-2011-Vestrand_WT-MNRAS.414.3537P};
\citealp{Li_L-2012-Liang_EW-ApJ.758.27L};
\citealp{Liang_EW-2013-Li_L-ApJ.774.13L};
\citealp{Wang_XG-2015-Zhang_B-ApJS.219.9W}).
For example, some GRBs exhibit late simultaneous bumps in their optical and X-ray afterglows
around the jet break, e.g., GRBs~100418A (\citealp{Marshall_FE-2011-Antonelli_LA-ApJ.727.132M,Laskar_T-2015-Berger_E-ApJ.814.1L}),
100901A (\citealp{Gorbovskoy_ES-2012-Lipunova_GV-MNRAS.421.1874G,Laskar_T-2015-Berger_E-ApJ.814.1L}),
120326A (\citealp{Melandri_A-2014-Virgili_FJ-A&A.572A.55M,Hou_SJ-2014-Geng_JJ-ApJ.785.113H,Laskar_T-2015-Berger_E-ApJ.814.1L})
, 120404A (\citealp{Guidorzi_C-2014-Mundell_CG-MNRAS.438.752G, Laskar_T-2015-Berger_E-ApJ.814.1L}).
Their optical afterglows seem to feature a long shallow decay before the late bumps.
An exemplar of these afterglows is GRB~120326A,
which exhibits simultaneous bumps in its optical and X-ray afterglows at around 35~ks after the burst trigger
(\citealp{Urata_Y-2014-Huang_K-ApJ.789.146U,Melandri_A-2014-Virgili_FJ-A&A.572A.55M}).
\cite{Hou_SJ-2014-Geng_JJ-ApJ.785.113H} propose
that a newborn millisecond pulsar with a strong wind is responsible for its late bumps.
\cite{Melandri_A-2014-Virgili_FJ-A&A.572A.55M} explore various physical scenarios
and conclude that the late bumps may be caused
by a refreshed external-forward shock or a geometrical effect.
By testing various physical processes,
\cite{Laskar_T-2015-Berger_E-ApJ.814.1L} argue that the energy injection into the external shock
is the most plausible mechanism for its late bump.
 They also argue that GRB~120404A, with optical and X-ray bumps at $t_{\rm obs}\backsimeq 1000$~s, can be explained in the same scenario.
\cite{Guidorzi_C-2014-Mundell_CG-MNRAS.438.752G} attribtuted
the optical/X-ray bumps of GRB~120404A to a decelerating jet viewed close to the jet edge,
combined with some early re-energization of the shock.
For GRB~100814A, \cite{Geng_JJ-2016-Wu_XF-ApJ.825.107G}
proposed  an ultrarelativistic electron-positron pair wind model to explain the late optical bumps at $t_{\rm obs}\backsimeq 10^4$~s.
\cite{Yu_YB-2015-Huang_YF-ApJ.805.88Y} invoked a magnetar with spin evolution to explain the afterglow emission.
\cite{Nardini_M-2014-Elliott_J-A&A.562A.29N} attributed the feature to the late-time activity of the central
engine in the observed afterglow emission.
Apart from the late bumps, we also note an another puzzling feature that
some bursts exhibit late simultaneous plateaux in their optical and X-ray afterglows around the jet break,
e.g., GRBs~120729A, 051111, and 070318 (\citealp{Huang_LY-2018-Wang_XG-ApJ.859.163H}).

The above mentioned late simultaneous bumps/plateaux are all directly followed by a steep decay,
which is very different from the decay behavior of afterglows preceding the late simultaneous bumps/plateaux
and thus is an intriguing phenomenon.
The models involved to explain the late simultaneous bumps/plateaux should also
explain the steep decay following the bumps/plateaux.
Actually, this feature (i.e., the simultaneous bumps/plateaux directly followed by a steep decay)
is strongly reminiscent of the afterglows in GRB~170817A,
which is formed in an off-axis observed external-forward shock propagating in an interstellar medium
(ISM, \citealp{Troja_E-2018-Piro_L-MNRAS.478L.18T,Lamb_GP-2019-Lyman_JD-ApJ.870L.15L,Troja_E-2019-vanEerten_H-MNRAS.tmp.2169T,Huang_BQ-2019-Lin_DB-MNRAS.487.3214H,Ren_J-2020-Lin_DB-ApJ.901L.26R}),
of which the mass density is constant.
The observations of GRB~170817A reveal an {\bf almost} achromatic bump followed by a steep decay,
which is very similar to the GRBs with late bumps mentioned above.
Therefore, we would like to suggest that the late simultaneous bumps in GRBs~100814A, 120326A, and 120404A may be formed in
an off-axis observed external-forward shock in a homogeneous medium.
We would like to point out that the mechanism proposed to explain GRB 170817A
has been recently extended to other bursts, e.g., GRBs~080503, 140903A, 150101B, and 160821B (\citealp{Fraija_N-2020-DeColle_F-ApJ.896.25F}).
Since there is a long shallow decay before the late optical bump in these GRBs,
which is different from that of GRB~170817A,
the shallow decay may indicate another kind of the circum-burst medium, e.g., the free stellar wind medium.

In this work, we study the emission of the external-forward shock
propagating in a free-to-shocked wind circum-burst environment,
where the mass density in the shocked wind is almost constant
(\citealp{Weaver_R-1977-McCray_R-ApJ.218.377W}; \citealp{Dyson_JE-1997-Williams_DA-pism.book.D}; \citealp{Chevalier_RA-2000-Li_ZY-ApJ.536.195C}).
The paper is originated as follows.
The dynamics and radiation of the external-forward shock are presented in Section~\ref{Sec:DYNAMICS OF THE EXTERNAL SHOCK}.
The theoretical afterglows are presented and discussed in Section~\ref{Sec:Results}.
As examples, we perform the fitting on the X-ray and optical afterglows of GRBs~120326A, 120404A, and 100814A,
and the results and discussions are shown in Section~\ref{Sec:Case Study}.
The conclusions are summarized in Section~\ref{Sec:Conclusion}.

\section{Dynamics and Radiation of the External-forward Shock}\label{Sec:DYNAMICS OF THE EXTERNAL SHOCK}
When a relativistic jet propagates in the circum-burst medium of a GRB,
an external shock would be developed and thus produces a long-term broadband afterglow emission
(\citealp{Sari_R-1998-Piran_T-ApJ.497L.17S}; \citealp{Meszaros_P-1999-Rees_MJ-MNRAS.306L.39M}; \citealp{Sari_R-1999-Piran_T-ApJ.517L.109S,Sari_R-1999-Piran_T-ApJ.520.641S}).
 To describe the jet structure and the dynamics of the external-forward shock,
we introduce a spherical coordinate ($R, \theta, \varphi$) with $R=0$ locating at the burst's central
engine and $\theta=0$ being along the jet axis.
The jet flow moving at the direction of ($\theta$, $\varphi$) is represented with ($\theta$, $\varphi$)-jet.
We assume the observer location at the direction of ($\theta_{\rm v}$, $\varphi_{\rm v}$)
with $\varphi_{\rm v}=0$ and $\theta_{\rm v}<\pi/2$.

In our calculations, the jet moving toward us
\footnote{The central engine of a GRB usually launches a pair of outflows,
i.e., a near-jet moving toward us and a counter-jet moving away from us.
In general, the emission from the counter-jet is negligible compared with the from the near-jet.
Then, we only consider the emission from the near-jet.}
is divided into $I\times L$ small patches along the $\theta$ and $\varphi$ directions in their linear space,
i.e.,
$[0,\;\delta \theta ],\;[\delta \theta ,\;2\delta \theta ],\;[2\delta \theta ,\;3\delta \theta ],\; \cdots,\;
[(i - 1)\delta \theta ,\;i\delta \theta ],\; \cdots
,\;[(I - 1)\delta \theta ,\;I\delta \theta ]$ with $\delta \theta  = {\theta _{\rm jet}}/I$
and $[0,\;\delta \varphi ],\;[\delta \varphi ,\;2\delta \varphi ],\;[2\delta \varphi ,\;3\delta \varphi ],\; \cdots,$ $[(l- 1)\delta \varphi ,\;l\delta \varphi ],\;\cdots,\;
[(L - 1)\delta \varphi ,\;L\delta \varphi ]$ with $\delta \varphi  = \pi /L$.
Here, $\theta_{\rm jet}$ is the opening angle of the jet.
The dynamics and the emission of the external-forward shock is estimated independently in each patch.
In addition, the following assumptions are adopted in estimating the dynamics and radiation of the external-forward shock:
(1) The sideways expansion are ignored
and the dynamics at each patch are assumed to be independent of other patches.
The amount of sideways expansion is a debated topic,
with numerical simulations suggesting a limited amount of spreading
(e.g., \citealp{Granot_J-2001-Miller_M-grba.conf.312G}; \citealp{Cannizzo_JK-2004-Gehrels_N-ApJ.601.380C}; \citealp{Zhang_W-2009-MacFadyen_A-ApJ.698.1261Z}; \citealp{vanEerten_H-2010-Zhang_W-ApJ.722.235V}; \citealp{vanEerten_HJ-2012-MacFadyen_AI-ApJ.751.155V}).
In addition, it is hard to consider the sideways expansion in the context of a structured jet
and the related afterglows fittings.
(2) The micro-physical parameters for synchrotron emission, e.g., $\epsilon_e$ and $\epsilon_B$, are set as constants, where $\epsilon_e$ and $\epsilon_B$ are the fractions of the shock
energy used to accelerate electrons and contributing to the magnetic energy, respectively.
(3) The energy injection in the external shock is not considered in our work.

\emph{Circum-burst environment:} For massive stars, associated with long GRBs, the stellar wind should be formed
and the structure of the wind bubble has been widely studied
(e.g., \citealp{Scalo_John-2001-Wheeler_JCraig-ApJ.562.664S};
 \citealp{GarciaSegura_G-1996-Langer_N-A&A.316.133G}; \citealp{Castor_J-1975-McCray_R-ApJ.200L.107C}).
Usually, it is accepted that
four regions, i.e., the free stellar wind, the shocked wind, the shocked ISM, and the unshocked ISM,
will be formed as the stellar wind interacting with the ISM (\citealp{Chevalier_RA-2000-Li_ZY-ApJ.536.195C}).
Since the latter two regions are too far to reach by the blast wave
(please see figure~1 of \citealp{Chevalier_RA-2000-Li_ZY-ApJ.536.195C}),
they are not considered here.
The mass density of the stellar wind medium is given by
${\rho _{_{\rm{free-wind}}}} = 5 \times {10^{11}}{A_*}{R^{ - 2}} \rm g \cdot c{m^{ - 1}}$ (\citealp{Chevalier_RA-2000-Li_ZY-ApJ.536.195C}),
where $A_*$ is a constant.
However, the mass density in the shocked wind depends on the uncertain physics of the heat conduction.
Heat conduction could be prevented by a magnetic field,
which is expected to be toroidal in this region (\citealp{Chevalier_RA-2004-Li_ZY-ApJ.606.369C}).
Under this assumption, and using the fact that the internal speed of sound
in this region is much higher than the expansion velocity,
the density in shocked wind is approximately $R$-independent
and equals to $\xi\rho_{\rm wind}(R_{\rm tr})$
with $\xi=4$ (\citealp{Weaver_R-1977-McCray_R-ApJ.218.377W}; \citealp{Dyson_JE-1997-Williams_DA-pism.book.D}; \citealp{Chevalier_RA-2000-Li_ZY-ApJ.536.195C}). Please see \cite{Peer_A-2006-Wijers_RAMJ-ApJ.643.1036P} for details.
Here, $R_{\rm tr}$, i.e., the free-wind boundary, is the transition radius from the free-wind
to the shocked-wind (\citealp{Kong_SW-2010-Wong_AYL-MNRAS.402.409K};
\citealp{Feng_Si-Yi-2011-Dai_Zi-Gao-RAA.11.1046F}; \citealp{Ramirez-Ruiz_Enrico-2001-Dray_LynnetteM-MNRAS.327.829R})
and depends on the age of the massive star wind and proper tracking of the shocked medium cooling.
In summary, the mass density of the  free-to-shocked wind circum-burst environment
is described as
\begin{eqnarray}\label{Eq:rho}
\rho(R)  = \left\{ {\begin{array}{*{20}{l}}
{ \rho_{_{\rm free-wind}}(R),}&{R < {R_{{\rm{tr}}}},}\\
{\rho_{_{\rm shocked-wind}}(R),\;\;}&{R \ge {R_{\rm tr}},}
\end{array}} \right.
\end{eqnarray}
where $\rho_{{\rm shocked-wind}}\equiv n_0m_{\rm p}{\rm cm^{-3}}=\xi\rho_{_{\rm free-wind}}(R_{\rm tr})$ and $m_{\rm p}$ is the proton mass.
One should note that the value of $\xi$ can be different from 4.
By fitting the afterglows,
\cite{Jin_ZP-2009-Xu_D-MNRAS.400.1829J} have obtained $\xi \approx 4.07$ for GRB~081109A
and
\cite{Feng_SY-2011-Dai_ZG-RAA.11.1046F} have found $\xi=2.1$, 2.0, and 2.5 for GRBs~080916C, 080916C, and 090926A, respectively.
For simplicity, $\xi=4$ is adopted in this paper.

\emph{Structured jet Description:}
In this paper,
we mainly focus on an axisymmetric structured jet propagating in a  free-to-shocked wind circum-burst environment.
Our structured jet at the radius $R_0$ is read as
\begin{eqnarray}\label{Eq:E_k}
\left\{ \begin{array}{l}
{E_{{\rm{k}},{\rm{iso}}}}(\theta,R_0)= {E_{{\rm{k}},{\rm{iso}},{\rm{on}}}}\exp [ - \frac{1}{2}{(\frac{\theta} {\theta _{\rm c}})^{\rm{2}}}],\\
\Gamma(\theta,R_0) = {\Gamma _0},
\end{array} \right.
\end{eqnarray}
where $E_{\rm k, iso}/(4\pi)$ is the kinetic energy per solid angle at $\theta$,
$E_{\rm k, iso, on}$ is corresponding to the value of $E_{\rm k, iso}$ at $\theta=0$,
and $\theta _{\rm c}$ is the characteristic angle of $E_{\rm k, iso}$.

\emph{Dynamics of the external-forward shock:}
For the ($\theta$, $\varphi$)-jet, the evolution of the bulk Lorentz factor $\Gamma(\theta,R)$ is described as
(\citealp{Huang_YF-1999-Dai_ZG-MNRAS.309.513H}; \citealp{Huang_YF-2000-Gou_LJ-ApJ.543.90H})
\begin{eqnarray}\label{eq:Gamma}
\frac{{d\Gamma(\theta, R)}}{{dR}} =  - \frac{{{\Gamma^2} - 1}}{{{M_{0}(\theta)} +  \epsilon m+ 2(1 -  \epsilon )\Gamma m}}\frac{dm(R)}{{dR}},
\end{eqnarray}
where $M_{0}(\theta)=E_{\rm k, iso}(\theta, R_0)/(4\pi\Gamma_0c^2)$ with $c$ being the light speed,
and $m$ is the sweep-up mass of the external shock from $R_0=10^{13}$~cm to $R$ per solid angle
and can be estimated with
\begin{eqnarray}\label{eq:m}
\frac{dm(R)}{dR}=\rho R^{2}.
\end{eqnarray}
In Equation~(\ref{eq:Gamma}), $\epsilon(\theta, R)=\epsilon_e\cdot\min[1, (\gamma'_{\rm e,m}/\gamma'_ {\rm e, c})^{p-2}]$
is the radiation efficiency of the external-forward shock,
where
$\gamma'_{\rm e, m} =\epsilon _e(p - 2){m_{\rm{p}}}\Gamma/[(p - 1){m_{\rm{e}}}]$
is the minimum Lorentz factor of the electrons accelerated in the shock,
$\gamma'_ {\rm e, c}=6\pi m_{\rm e} c/(\sigma_{\rm T}\Gamma {B'}^2 t_{\rm obs}^{\rm on})$ is
the efficient cooling Lorentz
factor of electrons (\citealp{Sari_R-1998-Piran_T-ApJ.497L.17S}).
Besides,
$B'(\theta, R)=(32\pi \rho \epsilon_B)^{1/2} \Gamma c$ is the magnetic field behind the shock,
$t_{{\rm obs}}^{\rm on}(\theta, R)=\int_{R_0}^{R} {(c - \upsilon)} {{dr}}/{c\upsilon}$
with $\upsilon(\theta, R)=c\beta$ and $\beta(\theta, R)=\sqrt {1 - 1/\Gamma^2}$,
and $m_{\rm e}$, $p$, and $\sigma_{\rm T}$
are the electron mass, the power-law index of accelerated electrons in the shock,
and the Thomson cross-section, respectively.

With Equations~(\ref{Eq:rho})-(\ref{eq:m}), one can estimate the value of $\Gamma$ at different $R$.
Then, the observed time for a photon from ($\theta$, $\varphi$)-jet can be estimated with
\begin{eqnarray}\label{Eq:tobs}
t_{\rm obs}(\theta, \varphi, \theta_{\rm v}, R)
=(1+z)\left [ \frac{{{R_0}}}{{2\Gamma _0^2c}}
+t_{\rm obs}^{\rm on}
+ \frac{R(1- \cos \Theta)}{c}\right ],
\end{eqnarray}
where $\cos \Theta  = (\sin {\theta}\cos {\varphi}{\rm{,}}\;\sin {\theta}\sin {\varphi},\;{\rm{cos}}{\theta}) \cdot (\sin {\theta _{\rm{v}}},\;0,\;\cos{\theta _{\rm{v}}}) =\sin {\theta}\cos {\varphi}\sin {\theta _{\rm{v}}} + \cos {\theta}\cos {\theta_{\rm{v}}}$,
and ${\Theta}$ is the angle between the direction of ($\theta$, $\varphi$) and the line of sight, i.e., ($\theta_{\rm v}$, $\varphi_{\rm v}$)
with $\varphi_{\rm v}=0$.
In Equation~(\ref{Eq:tobs}), the fisrt two terms in the right-hand side in the square bracket is the arrival time of photons
for an observer being in the direction of ($\theta_i$, $\varphi_l$).
Then, if the observer being in the direction of ($\theta_{\rm v}$, $\varphi_{\rm v}$) with $\varphi_{\rm v}=0$, the last term should be added.
For a given observer time $t_{\rm obs}$,
one can obtain the corresponding value of $R=R_{\rm obs}(\theta, \varphi, \theta_{\rm v})$ based on Equation~(\ref{Eq:tobs}).
The value of $R_{\rm obs}$ is the location of the external-forward shock
for the ($\theta$, $\varphi$)-jet observed at $t_{\rm obs}$
and is used to calculate the observed flux from the ($\theta$, $\varphi$)-jet.

\emph{Evolution of the electron energy spectrum:}
In the X-ray and optical bands,
the main radiation mechanism of the external-forward shock in GRBs
is the synchrotron radiation of the sweep-up electrons
(\citealp{Sari_R-1998-Piran_T-ApJ.497L.17S}; \citealp{Sari_R-1999-Piran_T-ApJ.517L.109S}).
We denote the instantaneous electron spectrum for the sweep-up electrons per solid angle
in the ($\theta$, $\varphi$)-jet as
$n'_{e}({\gamma'_{\rm e}},\theta, R)$,
where $n'_{e}d\gamma'_{\rm e}$ is the number of electrons in $[\gamma'_{\rm e}, \gamma'_{\rm e}+d\gamma'_{\rm e}]$
with $\gamma'_{\rm e}$ being the Lorentz factor of electrons.
The evolution of $n'_{\rm e}$ can be described as
\begin{eqnarray}\label{Eq:ne}
\frac{{\partial {n'_{\rm e}}}}{{\partial t'}} + \frac{\partial }{{\partial {\gamma'_{\rm e}}}}\left( {{{\dot \gamma '}_{\rm e}}{n'_{\rm e}}} \right) = Q',
\end{eqnarray}
where $\dot \gamma'_{\rm e}$ is the cooling rate of electrons and
$Q'$ is the injection rate of electrons from the shock, i.e.,
\begin{eqnarray}\label{Eq:Cooling}
{\dot \gamma'_{\rm e}}({\gamma'_{\rm e}},\theta, R) \equiv \frac{d\gamma'_{\rm e}}{dt'}
=  - \frac{{{\sigma _{\rm T}}{\gamma'_{\rm e}}^2{{B'}^2}}}{{6\pi {m_e}c}}
- \frac{2}{3}\frac{\gamma'_{\rm e}}{R}\frac{{dR}}{{dt'}}.
\end{eqnarray}
\begin{eqnarray}\label{Eq:Q}
Q'({\gamma'_{\rm e}},\theta, R)= \left\{ {\begin{array}{*{20}{c}}
{Q'_0{\gamma'_{\rm e}}^{ - p},}&{{\gamma'_{\rm e,m}}\leqslant{\gamma'_{\rm e}}\leqslant{\gamma'_{\rm e,{\rm{max}}}}},\\
{0,\;}&{\rm others,}
\end{array}} \right.
\end{eqnarray}
Here, $p\,(>2)$ is the power-law index,
${\gamma'_{\rm e,{\rm{max}}}} = \sqrt {9m_{\rm e}^2{c^4}/(8B'{q_e^3})}$
with $q_e$ being the electron charge (e.g., \citealp{Kumar_P-2012-Hernandez_RA-MNRAS.427L.40K}),
$Q'_0$ is obtained by solving $\int_{{\gamma'_{\rm e,m}}}^{{\gamma'_{\rm e,\max }}} Q'd{\gamma'_{\rm e}}  = ({R^2}\rho /{m_{\rm{p}}})dR/dt'$ with $dR/dt'=c\beta\Gamma$,
and $t'(\theta, R)$ is the time elapsed in the comoving frame of the blast wave.
The first and second terms in the right-hand side of Equation~(\ref{Eq:Cooling})
are respectively the synchrotron radiative cooling and adiabatic cooling of electrons,
and the inverse-Compton cooling is not considered here.
We solve Equation~(\ref{Eq:ne}) and $dR/dt'=c\beta\Gamma$ for $n'_{\rm e}({\gamma'_{\rm e}},\theta, R)$ at different $R$.
In our calculations, the fourth-order Runge-Kutta method is used.
In addition, an appropriate time step $\Delta {t'} < \min \left\{ {\left| {\Delta {\gamma'_{\rm e}}/{{\dot \gamma }_{\rm e}}} \right|} \right\}$ is adopted in our calculations,
where ${\Delta \gamma'_{\rm e}}$ is the width of our adopted energy grids for electrons
(e.g., see appendix A of \citealp{Geng_JJ-2018-Huang_YF-ApJS.234.3G}).

\emph{Observed flux calculation:}
With the obtained $n'_{\rm e}(\gamma'_{\rm e},\theta, R)$,
the spectral power of synchrotron radiation at a given frequency $\nu'$
can be described as
\begin{eqnarray}\label{EQ:SynPower}
P'(\nu',\theta, R)=\frac{\sqrt{3}q_{\rm e}^3B'}{m_{\rm e}c^2}\int_{0}^{\gamma'_{\max}}F\left({\nu'}/{\nu'_{\rm syn}}\right)n'_{\rm e}({\gamma'_{\rm e}},\theta, R)d\gamma'_{\rm e},
\end{eqnarray}
where $F(x)=x\int_{x}^{+\infty}K_{5/3}(k)dk$ with $K_{5/3}(k)$ being the modified Bessel function of 5/3 order
and $\nu'_{\rm syn}=3q_{\rm e}B'(\theta, R){\gamma'_{\rm e}}^2/(4\pi m_{\rm e}c)$.
Then, the observed flux density $\hat{f}_{\nu}(\theta, R)$
from the per solid angle of the jet flow in the direction of ($\theta$, $\varphi$) is
\begin{eqnarray}\label{EQ:ObsFlux}
{\hat{f}_{\nu}(\theta, \varphi, \theta_{\rm v},R)}=(1 + z)P'\left( {\nu \frac{{1 + z}}{D},\theta, R} \right){D^3}\frac{1}{{4\pi d_{\rm{L}}^2}},
\end{eqnarray}
where $\nu$ is the observed photon frequency,
$d_{\rm L}$ is the luminosity distance at the cosmological redshift $z$,
and $D(\theta, \varphi, \theta_{\rm v},R)= 1/[ \Gamma(1-\beta\cos\Theta)]$ is the Doppler factor of the ($\theta$, $\varphi$)-jet relative to the observer.
Then, the observed total flux density $f_{\nu}$ at $t_{\rm obs}$ is
\begin{eqnarray}\label{EQ:ObsFluxTotal}
f_{\nu} ({t_{{\rm{obs}}}}) = 2\sum\limits_{i = 1}^I {\sum\limits_{l = 1}^L {{{{\hat{f}_{\nu}}(\theta_i, \varphi_l, \theta_{\rm v}, R_{\rm obs})\sin \theta_i \delta \theta \delta \varphi}}} },
\end{eqnarray}
where $\theta_i=(i-0.5)\delta \theta$ and $\varphi_l=(l-0.5)\delta \varphi$.
The observed flux $F_{\rm XRT}$ in the X-ray band ($0.3-10$~keV) of the X-ray telescope (XRT) aboard \emph{Swift} satellite
 is calculated with
${F_{{\rm{XRT}}}} = \int_{{\nu _{_{0.3}}}}^{{\nu _{_{10}}}} {{f_\nu }({t_{{\rm{obs}}}})d\nu } $,
where $h\nu_{_{10}}=10$~keV and $h\nu_{_{0.3}}=0.3$~keV.

\section{Theoretical Afterglows }\label{Sec:Results}
In this paper, we mainly focused on the afterglows of GRBs in a  free-to-shocked wind circum-burst environment.
In Figure~{\MyFigA}, we shows the light-curves (solid lines) of the radiation from the external-forward shock
propagating in a  free-to-shocked wind circum-burst environment,
where $E_{\rm k,iso,on}=10^{52}$~erg, $\Gamma_0=250$,
$\theta_{\rm c}=5^\circ$, $\theta_{\rm jet}=8\theta_{\rm c}$,
$p=2.2$, $\epsilon_{e}=0.1$, $\epsilon_{B}=10^{-4}$, $A_{*}=0.01$, $R_{\rm {tr}}=10^{17} \rm {cm}$, and $z=1$
are adopted to calculate the dynamics and emission of the external-forward shock\footnote{
In general, the value of $A_{*}\sim 1 $ is assumed.
However, the value of $A_{*} < 1$ is generally reported in fitting the afterglows of GRBs,
e.g.,
$ A_{*} \backsimeq 0.005 - 0.032 $ for GRB~050319  (\citealp{Kamble_Atish-2007-Resmi_L-ApJ.664L.5K}),
$ A_{*} \backsimeq  0.02$ for GRB~081109A (\citealp{Jin_ZP-2009-Xu_D_MNRAS.400.1829J}),
$ A_{*} \backsimeq  0.2$ for GRB~160625B(\citealp{Fraija_N-2017-Veres_P-ApJ.848.15F}),
and $ A_{*} \backsimeq  0.06$ for GRB~190114C (\citealp{Fraija_N-2019-Dichiara_S-ApJ.879L.26F}).
Since the value of $A_{*} \sim 0.01$ is obtained from these fittings,
$A_{*} \sim 0.01$ is adopted to discuss the theoretical light-curves
for an external-forward shock propagating in a free-to-shocked wind circum-burst environment.}.
One can find that the XRT/optical light-curves (solid lines) from a low viewing angle
(e.g., $\theta_{\rm v}\lesssim \theta_{\rm c}$) can be decomposed into three phases:
\textcircled{\footnotesize{i}}
the free-wind-phase,
which appears in the early stage
and is formed during the external-forward shock propagating in the free-wind environment;
\textcircled{\footnotesize{ii}}
the shocked-wind-phase,
which appears in the middle section bridging the free-wind-phase to the post-jet-break-phase
and is formed during the external-forward shock propagating in the uniform shocked-wind environment;
\textcircled{\footnotesize{iii}}
the post-jet-break-phase,
which appears in the late stage as a steep decay
by reflecting the situation that the jet core is  fully visible for the observer,
i.e., $\Gamma(\theta_{\rm c}, R)\cdot \sin(\theta_{\rm c}+\theta_{\rm v})\lesssim 1$
\footnote{For a low value of $\theta_{\rm c}+\theta_{\rm v}$,
$\Gamma_{\rm c}\cdot \sin(\theta_{\rm c}+\theta_{\rm v})$ is reduced to
$\Gamma_{\rm c}\cdot (\theta_{\rm c}+\theta_{\rm v})$,
which is generally used in discussing the jet-break behavior.
For a high value of $\theta_{\rm c}+\theta_{\rm v}$, however, the jet-break appears in the situation with
$\Gamma_{\rm c}\cdot \sin(\theta_{\rm c}+\theta_{\rm v})\sim 1$ rather than
$\Gamma_{\rm c}\cdot (\theta_{\rm c}+\theta_{\rm v})\sim 1$.
}.
Here, $\Gamma(\theta_{\rm c}, R)$ is the bulk Lorentz factor of the jet flow
in the direction with $\theta=\theta_{\rm c}$,
$\theta_{\rm c}$ is the characteristic angle of $E_{\rm k, iso}$,
and the direction of ($\theta_{\rm v}$, $\varphi_{\rm v}$) with $\varphi_{\rm v}=0$ is the line of sight.

Figure~{\MyFigA} reveals that the transition from a free-wind to a homogeneous shocked-wind generally
witnesses the flatting/brightening of the light-curves.
By increasing the viewing angle,
the changes in light-curves of Figure~{\MyFigA} (solid lines) can be summarized as follows.
\textcircled{\footnotesize{1}}
Firstly, the free-wind-phase and the shocked-wind-phase become shallow while the post-jet-break phase remains its morphology.
\textcircled{\footnotesize{2}}
At a certain viewing angle, e.g., $\theta_{\rm v}\sim 4\theta_{\rm c}$,
the late simultaneous plateaux in the optical and X-ray afterglows appear.
It should be pointed out that these plateaux are all directly followed by a post-jet-break-phase,
which is similar to the late plateau found in the afterglows of
GRBs~120729A, 051111, and 070318 (\citealp{Huang_LY-2018-Wang_XG-ApJ.859.163H}).
\textcircled{\footnotesize{3}}
Further increasing the viewing angle,
the free-wind-phase turns into a rising pattern.
Meanwhile, the free-wind-phase and post-jet-break-phase remain their decaying morphology.
Then, the late simultaneous bumps in the optical and X-ray afterglows appear.
It should be pointed out that these late bumps are directly followed by a post-jet-break-phase
and preceded by a shallow decay,
which is very similar to the late simultaneous bumps found in some bursts,
e.g., GRBs~100418A, 100901A, 120326A, and 120404A.
The behaviors of \textcircled{\footnotesize{2}} and \textcircled{\footnotesize{3}}, i.e.,
the late simultaneous bumps/plateaux followed by a post-jet-break-phase in the optical and X-ray afterglows,
are the key findings of this work.
In Figure~{\MyFigB}, we also shows X-ray afterglows (solid lines)
of the external-forward shock in a free-to-shocked wind environment
for a jet with $\theta_{\rm jet}=\theta_{\rm c}$,
where the values of the other parameters are the same as those in Figure~{\MyFigA}.
One can find that the free-wind-phase in Figure~{\MyFigB}
is shallower than that in Figure~{\MyFigA} for the light-curves from large viewing angles.
Then, the opening angle of the jet may affect the light-curves before the late bump/plateau.
It is worthy to point out that
the only difference in producing
the light-curves of Figure~{\MyFigA} or {\MyFigB}
is the viewing angle $\theta_{\rm v}$.
Thus, the appearance of the late simultaneous bumps/plateaux in Figure~{\MyFigA} or {\MyFigB}
is related to the differences in the viewing angle rather than in the dynamics of the external shock.

For the structured jet adopted in Figures~{\MyFigA} and {\MyFigB},
the deceleration radius of the on-axis jet flow is in the free-wind medium,
where the deceleration radius of the ($0$, $0$)-jet is $1.17\times 10^{14}\rm cm$ based on Equation~(\ref{Eq:R_dec}).
Then, there is no broad late bumps in the afterglows for an on-axis observer.
In fact, the late simultaneous bumps/plateaux appear only in the cases with high viewing angle
for the free-to-shocked-wind-afterglows of Figure~{\MyFigA} or {\MyFigB}.
We would like to point out that
if the deceleration radius of the on-axis jet flow lies beyond or at around the free-wind boundary,
a broad bump preceded by a long plateau or shallow decay can appear in the afterglows
even in the cases with low viewing angle.
In Figure~{\MyFigC}, we shows the light-curves of the radiation from the external-forward shock propagating in a free-to-shocked wind environment,
where a structured jet with $E_{\rm k,iso,on}=10^{55}$erg, $\Gamma_0=100$,
$\theta_{\rm jet}=8\theta_{\rm c}$,
$p=2.2$, $\epsilon_{e}=0.1$, $\epsilon_{B}=10^{-4}$, $A_{*}=0.01$, $R_{\rm {tr}}=10^{17} \rm {cm}$,
and $\theta_{\rm c}=5^\circ$ (top panel) or $\theta_{\rm c}=0.5^\circ$ (bottom panel) is adopted.
For the given structured jet in Figure~{\MyFigC},
the deceleration radius of the ($0$, $0$)-jet in the free-wind medium
is $7.33\times 10^{17}\rm cm$, which is significantly larger than $R_{\rm tr}$.
Then, the deceleration radius of the on-axis jet flow would lie beyond the free-wind boundary.
In Figure~{\MyFigC}, late broad bumps preceded by a long plateau or shallow decay
indeed appear in the afterglows for the cases with low viewing angle.
In addition, the late broad bump is followed by a normal/steep decay
if a structured jet with high/low characteristic angle $\theta_{\rm c}$
(e.g, $\theta_{\rm c}=5^\circ$/$\theta_{\rm c}=0.5^\circ$) is adopted.

Now, we can conclude that
the late simultaneous broad bumps directly followed by a steep decay
and preceded by a long plateau or shallow decay
may sound the external-forward shock propagating in a free-to-shocked wind environment.
In addition, one of the following conditions should be satisfied:
\textcircled{\footnotesize{1}} the observer should have a high viewing angle with respect to the jet axis;
\textcircled{\footnotesize{2}} the structured jet should have a low characteristic angle
and the deceleration radius of the on-axis jet flow is at around or beyond the free-wind boundary.
The reason for the appearance of the broad bumps in condition \textcircled{\footnotesize{2}} or Figure~{\MyFigC} is clear.
However, the reason for the appearance of the bumps/plateaux in condition
\textcircled{\footnotesize{1}} or Figures~{\MyFigA} and {\MyFigB} is unclear.
Then, the following subsections, i.e., Subsection~\ref{Sec:Subsec1}-\ref{Sec:Subsec3},
are dedicated to discuss the late simultaneous bumps/plateau in Figures~{\MyFigA} and {\MyFigB}.

\subsection{Understanding the Appearance of Late bump/plateau}\label{Sec:Subsec1}
The appearance of the bump/plateau in Figure~{\MyFigA} or {\MyFigB} can be understood as follows.
In Figures~{\MyFigA} and {\MyFigB}, we also plot the afterglows formed in a pure free-wind medium with $\rho=5\times 10^{11} A_{*}R^{-k}\rm g \cdot c{m^{ - 3}}$ (dotted lines)
and those formed in a pure shocked-wind medium with $\rho=n_{0}m_{p}{\rm cm^{-3}}$ (dashed lines) for some viewing angles.
(Hereafter, the afterglow formed in a  free/shocked/free-to-shocked wind environment is represented with ``free/shocked/free-to-shocked-wind-afterglow''.)
One can find that
the light-curves of the shocked-wind-phase and the post-jet-break-phase
are almost the same as those of the shocked-wind-afterglow in the same time interval.
Then, the afterglow formed in a free-to-shocked wind environment
can be approximately estimated by superposition of the free-wind-afterglow
and the shocked-wind-afterglow from different viewing angle.
By increasing the viewing angle,
the decay of the free-wind-afterglows and the shocked-wind-afterglows become shallow.
Then, the free/shocked-wind-phase in the free-to-shocked-wind-afterglows become shallow correspondingly.
Similar to the afterglow of GRB~170817A,
a late bump followed by a post-jet-break phase is presented in the shocked-wind-afterglow
while it is viewed at a large viewing angle, e.g., $\theta_{\rm v}\gtrsim 4 \theta_{\rm c}$.
However, the free-wind-afterglow remains its decaying pattern before the late bump of the shocked-wind-afterglow.
Then, a late bump/plateau followed by a post-jet-break-phase and preceding by a plateau/shallow decay
emerges by superposition of the
free-wind-afterglow and the shocked-wind-afterglow.
This is the phenomenological reason for the appearance of the late bump/plateau.

In this paragraph, we discuss the morphology of the free/shocked-wind-afterglows
for high-viewing-angle observers.
For the sake of discussion, we first study the morphology of the free/shocked-wind-afterglows
for a top-hat jet.
In Appendix~\ref{Sec_LC_TopHat}, the analytical results of the free/shocked-wind afterglow for a top-hat jet are derived
and summarized in Table~{\MyTabB}.
It can be found that high-viewing-angle observers generally witnesses
a rising pattern before the post-jet-break-phase for the afterglows from a top-hat jet.
Besides, the rise of the free-wind-afterglow is gentle compared with the rise of the shocked-wind-afterglow,
of which the rising power-law index $-\alpha$ is larger than $4$.
This behavior can be found in Figure~{\MyFigB},
where a Gaussian structured jet with $\theta_{\rm jet}=\theta_{\rm c}$ describes a top-hat jet approximately.
For high-viewing-angle observers (e.g., $\theta_{\rm v}>\theta_{\rm c}$),
Figure~{\MyFigB} reveals that the rising power-law index $-\alpha$ is around $0.5$
in the free-wind afterglows and is around $4$ in the shocked-wind afterglows.
Correspondingly, the free-wind afterglows appear as an extremely wide bump with a slow rise,
and the shocked-wind afterglows appear as a narrow bump with a fast rise.
These results are consistent with those summarized in Table~{\MyTabB}.
In Figure~{\MyFigD}, we shows the free-to-shocked-wind-afterglows for a Gaussian structured jet with different $\theta_{\rm jet}$, where $\theta_{\rm v}=4\theta_{\rm c}$ is set.
One can find that the emission of the jet core mainly makes its contribution to the afterglow around the jet break for high-viewing-angle observers,
and the emission of the jet flow being close to the line of sight
dominates the early phase of afterglow.
Figure~{\MyFigD} also reveals that the superposition of the emission from the jet flow
being close to the line of sight
to the emission of the jet core generally slows down the rise of the early afterglows,
and even changes the rising pattern into the decaying pattern for the early afterglows.
Then, there is a long plateau or shallow decay before the late bump in our theoretical light-curves.

\subsection{Characteristic Transition Time}
The free-to-shocked afterglows have two characteristic transition time:
the transition from free-wind phase to the shocked-wind-phase (named as ``the first-transition'' in this paper);
the transition from the shocked-wind-phase to the post-jet-break-phase
(named as ``the second-transition'' in this paper).

The first-transition may be related to the observed time of the jet flow
crossing $R_{\rm tr}$.
This can be found by comparing the light-curves in Figure~{\MyFigD}
for the situations with $\xi=4$ (solid lines), $\xi=16$ (dotted lines), $\xi=64$ (dashed lines),
where the values of $A_*$ and $n_0$ are respectively the same in different situations,
and the value of $R_{\rm tr}$ is changed in order to obtain different $\xi=\rho_{{\rm shocked-wind}}/\rho_{_{\rm free-wind}}(R_{\rm tr})$.
The increase of $\xi$ in Figure~{\MyFigD} is related to the increase of $R_{\rm tr}$.
Figure~{\MyFigD} reveals that a higher value of $R_{\rm tr}$ adopted,
a later of first-transition appears.
In Figure~{\MyFigA},
we plot the observed times of the ($\theta_{\rm v}$, $\varphi_{\rm v}$)-jet
and ($0$, $0$)-jet\footnote{The ($0$, $0$)-jet represents the jet flow along the jet axis.
For a high-viewing-angle observer, the observed time of the ($0$, $0$)-jet crossing $R_{\rm tr}$
can be analytically estimated with Equation~(\ref{Eq:AP_Dynamics_Off-axis}),
i.e., $t_{{\rm{obs}}}(0, 0, \theta_{\rm v}, R_{\rm tr})= (1 + z)(1 - \cos {\theta _{\rm{v}}}){R_{{\rm{tr}}}}/c$,
which is consistent with that obtained based on the numerical method
(i.e., ``{\LARGE $\circ$}'' symbols in Figure~{\MyFigA}).
}
crossing $R_{\rm tr}$ with ``{\Large $\bullet$}'' and ``{\LARGE $\circ$}'' symbols, respectively.
(Hereafter, the symbols and lines with a same color in a same figure corresponds to the situation with a same viewing angle.)
One can find that the first-transition occurs during the
($\theta_{\rm v}$, $\varphi_{\rm v}$)-jet crossing $R_{\rm tr}$ for low-viewing-angle observers.
For high-viewing-angle observers,
however, the time for the ($\theta_{\rm v}$, $\varphi_{\rm v}$)-jet crossing $R_{\rm tr}$
is significantly larger than the first-transition time.
This implies that the free-wind-phase is shaped by
the emission of the jet flow at around ($\theta$, $\varphi$)=($\theta_{\rm v}$, $\varphi_{\rm v}$)
only in the situations with low viewing angle.
Figure~{\MyFigA} also reveals that the observed time of the ($0$, $0$)-jet crossing $R_{\rm tr}$, i.e., the ``{\LARGE $\circ$}'' symbols, is generally larger than the first-transition time.
Then, we can believe that the first-transition is not associated
with the dynamics of the ($\theta_{\rm v}$, $\varphi_{\rm v}$)-jet or ($0$, $0$)-jet for high-viewing-angle situations.
In Figure~{\MyFigD}, the first-transition occurs later by decreasing $\theta_{\rm jet}$.
Then, one can believe that the first-transition should be related to the dynamics of the jet flow being in the direction between ($\theta_{\rm v}$, $\varphi_{\rm v}$) and (0, $0$) directions.
We have found that the observed times of
the ($2\theta_{\rm c}$, $\varphi_{\rm v}$)-jet and ($3\theta_{\rm c}$, $\varphi_{\rm v}$)-jet crossing $R_{\rm tr}$ are at around the first-transition for the situations with $\theta_{\rm v}=4\theta_{\rm c}$ and $\theta_{\rm v}=8\theta_{\rm c}$, respectively.

The second-transition from the shocked-wind-phase to the post-jet-break-phase
may be related to the following three observed times\footnote{
If $\Gamma_0\gg 1/\sin ({\theta _{\rm{v}}} - {\theta _{\rm{c}}})$ is satisfied,
the observed time of \textcircled{\scriptsize{i}}, \textcircled{\scriptsize{ii}},
and \textcircled{\scriptsize{iii}} can be analytically estimated with
${\left[ {\sin ({\theta _{\rm{v}}} - {\theta _{\rm{c}}})} \right]^{ - \frac{{\varepsilon  + 2s - 8}}{{3 - s}}}}\Gamma _0^{ - \frac{{\varepsilon  + 2s - 8}}{{3 - s}}}\zeta {t_{{\rm{dec}}}}$,
${\left[ {\sin {\theta _{\rm{v}}}} \right]^{ - \frac{{\varepsilon  + 2s - 8}}{{3 - s}}}}\Gamma _0^{ - \frac{{\varepsilon  + 2s - 8}}{{3 - s}}}\zeta {t_{{\rm{dec}}}}$,
and
${\left[ {\sin ({\theta _{\rm{v}}} + {\theta _{\rm{c}}})} \right]^{ - \frac{{\varepsilon  + 2s - 8}}{{3 - s}}}}\Gamma _0^{ - \frac{{\varepsilon  + 2s - 8}}{{3 - s}}}\zeta {t_{{\rm{dec}}}}$
based on Equation~(\ref{Eq:AP_Dynamics_On-axis}), respectively.
The analytical results of above three observed times are consistent with these estimated based on numerical method, i.e., the values shown with ``{\Large $\triangleright$}'',
``{\Large $\diamond$}'', and ``{\Large $\triangleleft$}'' symbols in Figures~{\MyFigA} and {\MyFigB}.}:
\textcircled{\footnotesize{i}} the near core-edge, i.e., ($\theta_{\rm c}$, $\varphi_{\rm v}$)-jet, begins to be visible for the observer, i.e., $\Gamma(\theta_{\rm c}, R)\sin(\theta_{\rm v}-\theta_{\rm c})\backsimeq 1$;
\textcircled{\footnotesize{ii}} the jet axis, i.e., ($0$, $0$)-jet, begins to be visible for the observer, i.e., $\Gamma(0,R)\sin(\theta_{\rm v})\backsimeq 1$;
\textcircled{\footnotesize{iii}}
the far core-edge, i.e., ($\theta_{\rm c}$, $\varphi_{\rm v}+\pi$)-jet, begins to be visible for the observer, i.e., $\Gamma(\theta_{\rm c}, R)\sin(\theta_{\rm v}+\theta_{\rm c})\backsimeq 1$.
It is worthy to point out that when the far core-edge becomes visible,
the afterglow fully enters into the post-jet-break phase
and appears as a steep decay.
In Figures~{\MyFigA} and {\MyFigB},
the observed times of \textcircled{\footnotesize{i}}, \textcircled{\footnotesize{ii}},
and \textcircled{\footnotesize{iii}} are indicated with ``{\Large $\triangleright$}'',
``{\Large $\diamond$}'', and ``{\Large $\triangleleft$}'' symbols, respectively.
It can be found that
the second-transition begins
at around ``{\Large $\triangleright$}''
and ends at ``{\Large $\triangleleft$}'' for an observer with low viewing angle.
In addition, the ``{\Large $\triangleright$}'' symbols are generally at around the end of the plateau or the peak of the bump if the late bump/plateau appears in the afterglows.
For high-viewing-angle observers, the shocked-wind-phase generally appears as a plateau or fast rise.
Then, the second-transition begins at the end of the plateau or fast rise,
i.e., the observed time of the ``{\Large $\triangleright$}'' symbols, for high-viewing-angle observers.
In summary, the second-transition begins
at around ``{\Large $\triangleright$}''
and ends at ``{\Large $\triangleleft$}''.

\subsection{Behaviors of the Late Bump/Plateau}\label{Sec:Subsec3}
We note that the energy injection into the external shock
is always used to explain the bump/plateau in the afterglows.
In the energy injection scenario,
it is worthy to point out that the light-curve of the afterglow
shaped by the energy injection is achromatic\footnote{
The energy injection model is always used to explain the simultaneous bumps/plateaux in the afterglows.
In this scenario, the rise of the bumps is generally shaped by the energy injection and thus is achromatic.
However, the decay of bumps in different bands may not follow a same morphology since the decay phase of bumps is generally not shaped by the energy injection.
}.
However, we find that the free-to-shocked-wind-afterglows from an off-axis observed structured jet
does not present an absolutely achromatic bump/plateau.
In Figure~{\MyFigE}, we shows the afterglows observed in the radio-band ($\nu = 10 \rm {GHz}$, dotted lines), $R$-band (dashed lines), and XRT band (solid lines) for the situation with $\theta_{\rm v}=2\theta_{\rm c}$ (blue lines), $3\theta_{\rm c}$ (orange lines), $4\theta_{\rm c}$ (magenta lines) or $8\theta_{\rm c}$ (violet lines).
One can find that
the bumps/plateaux in the radio, optical, and X-ray afterglows
are simultaneous but not absolutely achromatic, especially for the situations with low viewing angle.
This behavior is different from that of the bump/plateau formed in the energy injection scenario.
The different behavior of simultaneous bumps/plateaux in the afterglows
from the energy injection scenario and those from our scenario
(i.e., an off-axis observed structured jet propagating in a free-to-shocked wind environment)
can be understood as follows.
The energy injection scenario and our scenario
may be not fundamentally different in shaping the simultaneous bumps/plateaux
because both scenarios try to account for light-curve morphology by varying the kinetic energy of the visible jet flow.
In the energy injection scenario, the kinetic energy varies in the lab and observer time
owing to the energy injection into the external shock.
In our scenario,
the kinetic energy of the visible jet flow implicitly varies with observer time
since the observer sees an increasing region of the structured jet owing to the deceleration of the jet.
However, the observed flux in our scenario
involves the emission from all of visible region of the structured jet
rather than that only from the region being just right visible.
Then, the appearance of bump or plateau may be simultaneous
but not absolutely achromatic in different observed bands.
Since our obtained bumps/plateaux appear simultaneous in the optical and X-ray afterglows,
our scenario could not fully account for chromatic bumps/plateaux.


In Figure~{\MyFigF}, we show the free/shocked/free-to-shocked-wind-afterglows
for a jet with $\theta_{\rm c}=0.5^\circ$,
where the values of other parameters are the same as those in Figure~{\MyFigA}.
One can find that the width of late bump/plateau in Figure~{\MyFigF}
is narrower than that in Figure~{\MyFigA}.
It reveals that the characteristic angle $\theta_{\rm c}$ of the structured jet may
affect the width of the late bump/plateau for the cases with a same $\theta_{\rm v}/\theta_{\rm c}$.
The reason is presented as follows.
In Figures~{\MyFigA} and {\MyFigB},
the peak time of bump or the end of plateau
is close to the beginning of the post-jet-break-phase.
In addition, the free-wind boundary is reached well before the post-jet-break-phase
for an on-axis observer (i.e., $\theta_{\rm v}=0$) and thus for an off-axis observer.
Then, one can find a broad bump/plateau for an observer with high viewing angle.
In Figure~{\MyFigF},
one can find that
the  free-wind boundary is reached during the post-jet-break-phase
for an observer with low viewing angle (e.g., $\theta_{\rm v}\lesssim 2\theta_{\rm c}$),
and is reached just before the post-jet-break-phase
for an observer with high viewing angle (e.g., $\theta_{\rm v}\gtrsim 4\theta_{\rm c}$).
Then, one can find a narrow bump/plateau.
By summary up the light-curves in Figures~{\MyFigA}, {\MyFigB}, and {\MyFigF},
one can conclude that
the late simultaneous broad bumps (e.g., GRB~120326A) only appear in the situation
that the free-wind boundary is reached well before the jet core becomes fully visible
for an observer with high viewing angle.

\section{Case Study} \label{Sec:Case Study}
As examples, we perform the fitting on the X-ray and
optical afterglows of GRBs~120326A, 120404A, and
100814A.
The light-curves from prompt emission to the late afterglow of these bursts are shown in Figure~{\MyFigE},
where the late bumps directly followed by a post-jet-break-phase
and preceded by a plateau or shallow decay
can be easy found in these light-curves.
\begin{itemize}
\item
GRBs~120326A is a long GRB at redshift $z = 1.798$ with unusual X-ray and optical afterglows.
The late simultaneous bumps followed by a post-jet-break-phase in the optical and X-ray afterglows
can be easily found at around $t_{\rm obs}\sim 4\times 10^{4}$~s.
Some authors proposed that the energy injection model may be responsible to this late bump
(e.g., \citealp{Melandri_A-2014-Virgili_FJ-A&A.572A.55M,Hou_SJ-2014-Geng_JJ-ApJ.785.113H,Laskar_T-2015-Berger_E-ApJ.814.1L}).
Since the observed bumps is very similar to those found in Figures~{\MyFigA}-{\MyFigC},
we would like to model the late simultaneous bumps
based on an off-axis observed external-forward shock in a free-to-shocked wind environment.
The XRT data at $t_{\rm obs}\gtrsim 3\times10^3$~s and the optical data at $t_{\rm obs}\gtrsim 10^2$~s are used in our fitting.

\item
GRB~120404A, with redshift $z = 2.876$, is a long GRB with
a significant rebrightening in the optical and
near-infrared bands at $t_{\rm obs}\sim 2\times 10^3$~s.
The X-ray observations around this time also show a bump
although the data are sparse owing to the orbital gap of \emph{Swift} (\citealp{Guidorzi_C-2014-Mundell_CG-MNRAS.438.752G}).
Interestingly, the bumps are all directly followed by a post-jet-break-phase.
\cite{Laskar_T-2015-Berger_E-ApJ.814.1L} proposed that
the energy injection model may be responsible to this late bump.
However, we would like to model these two late simultaneous bumps
based on an off-axis observed external-forward shock in a  free-to-shocked wind environment
since these bumps are similar to those in Figures~{\MyFigA}-{\MyFigC}.
The XRT data at $t_{\rm obs}>550$~s and the optical data at $t_{\rm obs}\gtrsim200$~s are used for our fitting.

\item
GRB~100814A is a long GRB with the redshift $z = 1.44$.
The optical rebrightening appears at $10^5$~s.
\cite{DePasquale_Massimiliano-2015-Kuin_NPM-MNRAS.449.1024} presented a broadband observations of GRB~100814A
and they attribute the late optical rebrightening to a long-lived external-reverse shock and external-forward shock.
Besides, \cite{Geng_JJ-2016-Wu_XF-ApJ.825.107G} used
an ultrarelativistic $e^+e^-$ wind injection model
to explain the late rebrightening.
It should be noted that
the optical/X-ray afterglows all turn into the steep decay
after the peak time of the optical bump.
Then, we would like to believe that
there may be a simultaneous late bumps in the optical and X-ray afterglows
and the rise of the X-ray bump is covered by the late central engine activities.
The XRT data at $t_{\rm obs}\gtrsim 9\times10^4$~s and
the optical data at $t_{\rm obs}\gtrsim 10^3$~s are also modelled as the emission
from an off-axis observed external-forward shock in a free-to-shocked wind environment.
\end{itemize}

Our fitting is performed based on
the Markov Chain Monte Carlo (MCMC) method to produce
posterior predictions for the model parameters.
MCMC method is widely used in finding a best set of parameters for a specified model,
e.g., GRB~080413B (\citealp{Geng_JJ-2016-Wu_XF-ApJ.825.107G}); GRBs~100418A, 100901A, 120326A, and 120404A (\citealp{Laskar_T-2015-Berger_E-ApJ.814.1L}).
In our work, fitting the afterglows of GRBs~120326A, 120404A, and 100814A with the MCMC method
is to test whether or not the late simultaneous bumps followed by a post-jet-break-phase
can be explained with an off-axis observed external-forward shock in a  free-to-shocked wind environment.
The posterior probability density functions for the physical parameters,
i.e.,
$E_{\rm k,iso, on}$, $\Gamma_0$, $\theta_{\rm c}$, $\theta_{\rm jet}/\theta_{\rm c}$,
$\theta_{\rm v}/\theta_{\rm c}$, $p$, $\epsilon_{e}$, $\epsilon_{B}$, $A_{*}$, and $R_{\rm tr}$,
are presented in Figure~{\MyFigH},
 where only the fitting result of GRB~120326A is shown as an example.
The optimal result from MCMC fitting is shown in Figure~{\MyFigG} with blue line (XRT) and red line (optical),
and the obtained parameters at the $1\sigma$ confidence level are reported in Table~{\MyTabA},
 where the values of the transition radius $R_{\rm {tr}}$
(i.e., $1.05 \times 10^{17}$cm, $6.31 \times 10^{16}$cm, and $3.98\times 10^{17}$cm for GRBs~120326A, 120404A and 100814A, respectively)
are consistent with those found in other bursts (e.g., \citealp{Kong_SW-2010-Wong_AYL-MNRAS.402.409K};
\citealp{Feng_Si-Yi-2011-Dai_Zi-Gao-RAA.11.1046F};
\citealp{Ramirez-Ruiz_Enrico-2001-Dray_LynnetteM-MNRAS.327.829R};
\citealp{Li_L-2020-Wang_XG-ApJ.900.176L}).
It can be found that both the X-ray afterglow
and the optical afterglow of these bumps can be well modelled
with an off-axis observed external-forward shock in a  free-to-shocked wind environment.
\begin{enumerate}
\item
 The fitting results reported in Table~{\MyTabA} reveals that
the on-axis observed isotropic kinetic energy $E_{\rm k,iso,on}$
of these bursts (i.e., $\sim 5\times 10^{54}\rm erg$, $6\times 10^{53}\rm erg$, and $8\times 10^{54}\rm erg$ for GRBs~120326A, 120404A, and 100814A, respectively)
are significantly high
but consistent with those found in other bursts (see the figure~10 in \citealp{Racusin_JL-2011-Oates_SR-ApJ.738.138R} and the table~1 in \citealp{Yi_SX-2017-Lei_WH-JHEAp.13.1Y}).
The total kinetic energy of the structured jet is also estimated
and shown in the last row of Table~{\MyTabA}.
Compared with the beaming corrected kinetic energy reported in the table~2 of \cite{Yi_SX-2017-Lei_WH-JHEAp.13.1Y},
the total kinetic energy of our obtained structured jet
is also consistent with those found in other bursts.
However, it should be noted that the on-axis observed luminosity of GRBs~120326A and 100814A may be significantly high since the value of $\theta_{\rm v}\sim 4\theta_{\rm c}$ is obtained from our MCMC fittings.
According to the structure of the jet,
the on-axis observed luminosity of these two bursts may be around $10^{54}\rm erg\cdot s^{-1}$,
which lies in the high-luminosity range of GRBs.
Here, the time-averaged luminosity of GRBs~120326A and 100814A are estimated to be around
$\sim 2\times 10^{51}\rm erg\cdot s^{-1}$ (\citealp{Melandri_A-2014-Virgili_FJ-A&A.572A.55M})
and $10^{51}\rm erg\cdot s^{-1}$ (\citealp{Nardini_M-2014-Elliott_J-A&A.562A.29N}), respectively.

\item
The theoretical light-curves does not well fit
the transition behavior from the wind-phase to the ISM-like phase in the afterglows of GRB~120326A.
It may imply that the density jump factor $\xi$ from the wind medium to the ISM-like medium may be less than 4,
i.e., $\xi<4$.
Figure~{\MyFigA} reveals that the external-forward shock propagating in a shocked-wind environment can yield a plateau before the late simultaneous bumps in the case with $\theta_{\rm v}=4\theta_{\rm c}$.
Then, we try to fit the afterglows of GRB~120326A with
the emission of the external-forward shock propagating
in a homogeneous medium,
where the priors of $\log_{10}(E_{\rm k,iso, on}/\rm erg)$, $\log_{10}\Gamma_0$, $\theta_{\rm c}$, $\theta_{\rm jet}/\theta_{\rm c}$,
$\theta_{\rm v}/\theta_{\rm c}$, $p$, $\log_{10}\epsilon_{e}$, $\log_{10}\epsilon_{B}$, and $\log_{10}n_0$
are set as uniform distribution in the range of
(52, 55), (1.5, 3.0), (0.3, 8.0), (2.5, 4.5), (0.0, 8.0), (2.1, 2.9), (-3.2, -0.5), (-6.5, -2.0), and (0.0, 1.7) in our MCMC fittings, respectively.
However, no well fitting result is found.
The fitting result with minimum reduced $\chi^2$ is shown with dashed lines in the top panel of Figure~{\MyFigG}. It can be found that the early plateau of GRB~120326A could not be well fitted
based on the emission of the external-forward shock in a homogeneous medium.

\item
There seems to be a plateau at $t_{\rm obs}\sim 600$~s in the optical afterglow of GRB~100814A,
which may indicate an energy injection into the external shock.
If so, our model could not well describe the afterglows of GRB~100814A in its early phase.
This may be the reason for the deviation of our theoretical results
relative to the observations in the early phase.
 In addition, the X-ray and optical afterglows of GRB~100814A show a chromatic behavior,
which could not be explained only with our scenario (i.e., an off-axis observed structured jet propagating into a free-to-shocked wind environment)
since the appearance of the late simultaneous bump/plateau in our scenario is quasi-achromatic.
Other emission rather than that of the external-forward shock may contribute to the long X-ray shallow decay.

\item
For the afterglows of GRB~120404, the fitting result shows that the line of sight
lies within the jet core, i.e., $\theta_{\rm v}=0.68\theta_{\rm c}$.
However, the late simultaneous bumps do not appear in the free-to-shocked-wind-afterglows of Figures~{\MyFigA} and {\MyFigB} for $\theta_{\rm v}<\theta_{\rm c}$.
Then, the appearance of the bumps in GRB~120404 should be related to
the condition \textcircled{\footnotesize{2}}, i.e.,
the structured jet has a low characteristic angle
and the deceleration radius of the on-axis jet flow is at around or beyond the free-wind boundary.
Based on the fitting result of this burst, the deceleration radius of
the ($0$, $0$)-jet can be estimated and is $2.43\times 10^{16}$~cm,
which is at around the transition radius $R_{\rm tr}=6.31 \times 10^{16}$~cm of this burst.
Then, the fitting result is consistent with the condition \textcircled{\footnotesize{2}}.

\end{enumerate}

\section{Conclusion}\label{Sec:Conclusion}
In this paper, the emission of the external-forward shock in a free-to-shocked wind
circum-burst environment is studied.
We mainly focus on the light-curves of the afterglows in the late phase.
A late  simultaneous broad bumps/plateaux followed by a post-jet-break-phase
and preceded by a plateau or shallow decay
is found in the obtained afterglows with one of the following conditions:
\textcircled{\footnotesize{1}} the observer should have a high viewing angle with respect to the jet axis;
\textcircled{\footnotesize{2}} the structured jet should have a low characteristic angle
and the deceleration radius of the on-axis jet flow is at around or beyond the free-wind boundary.
Our obtained bump/plateau in the afterglows is similar to those found in the afterglows of some bursts,
e.g., the late bump found in GRBs~100418A, 100901A, 120326A, 120404A
(see \citealp{Laskar_T-2015-Berger_E-ApJ.814.1L}),
080413B, 091029, 100814A, and 090426 (see \citealp{Geng_JJ-2016-Wu_XF-ApJ.825.107G}),
and the late plateau found in GRBs~120729A, 051111, and 070318 (see \citealp{Huang_LY-2018-Wang_XG-ApJ.859.163H}).
As examples,
we perform the fitting on the optical and X-ray afterglows of GRB~100814A, GRB~120326A and GRB~120404A
based on the Markov Chain Monte Carlo method.
It is found that the external-forward shock propagating in a free-to-shocked wind
circum-burst environment
for an off-axis observer can well explain the late phase of the optical and X-ray afterglows in these bursts.
 Since our obtained bumps/plateaux appear simultaneous in the optical and X-ray afterglows,
our scenario could not fully account for chromatic bumps/plateaux, e.g., GRB~100814A.
Then, we suggest that a late broad bump/plateau followed by a post-jet-break-phase
and preceded by a plateau or shallow decay
may sound the transition of the circum-burst environment from a free-wind to
a homogeneous shocked-wind medium.

In this work, a Gaussian structured jet is adopted to fit the late phase of afterglows in GRBs~100814A, 120326A and 120404A. However, we would like to point out that our conclusion about the circum-burst environment (i.e., the free-to-shocked wind environment) in these bursts may be robust.
The reasons are shown as follows.
Firstly, the late simultaneous bumps found in these bursts
are very similar to those in GRB~170817A,
which is formed in an off-axis observed external-forward shock propagating in an interstellar medium
(\citealp{Troja_E-2018-Piro_L-MNRAS.478L.18T,Lamb_GP-2019-Lyman_JD-ApJ.870L.15L,Troja_E-2019-vanEerten_H-MNRAS.tmp.2169T,Huang_BQ-2019-Lin_DB-MNRAS.487.3214H,Ren_J-2020-Lin_DB-ApJ.901L.26R}).
This implies that the late bumps in GRBs~100814A, 120326A, and 120404A may be formed in
an off-axis observed external-forward shock in a homogeneous medium.
Interestingly, there seems to be no plateau/shallow decay in the early phase of GRB~170817A's afterglows
(see the figure~1 of \citealp{Ren_J-2020-Lin_DB-ApJ.901L.26R}, especially the X-ray afterglow in $t_{\rm obs}\sim2-100$~day).
This behavior may also appear in the afterglows of GRBs~120326A and 120404A
if their late simultaneous bumps appear with the same mechanism as that in GRB~170817A.
In this paradigm, the plateau/shallow decay preceding the bumps in these two bursts
may require an another explanation
and the external-forward shock in the free-wind environment may be a natural candidate,
especially for long GRBs.
Secondly, the external-forward shock with other kinds of structured jet are also studied.
For example,
\cite{Huang_BQ-2019-Lin_DB-MNRAS.487.3214H} has studied the jet structure owing to the jet precession and the corresponding afterglows in a homogeneous environment.
Very complex structured jets have been found.
\cite{Granot_J-2018-Gill_R-MNRAS.481.1597G} studies
the afterglows from 2D relativistic hydrodynamic simulations of a GRB jet propagating in a homogeneous medium.
It seems to be the case that the late simultaneous bumps generally appear without preceding by a plateau/shallow decay in the afterglows for a structured jet propagating in a homogeneous medium
(e.g., the figures~3 and 5 of \citealp{Huang_BQ-2019-Lin_DB-MNRAS.487.3214H}, the figures~6 and 7 of \citealp{Granot_J-2018-Gill_R-MNRAS.481.1597G}).
However, our work could not rule out the scenario
that other kinds of structured jet propagating only in a homogeneous medium could explain the observed late simultaneous bumps
directly followed by a steep decay and preceded by a plateau/shallow decay in these burst.
Our scenario,
i.e., an off-axis observed structured jet propagating into a free-to-shocked wind environment,
only provides a possible way to decipher these bumps.

\acknowledgments
We thank the anonymous referee for beneficial suggestions that improved the paper.
This work is supported by the National Natural Science Foundation of China
(grant Nos.11773007, 11503011, 11673006, U1938201, U1731239) and the Guangxi Science Foundation
(grant Nos. 2018GXNSFFA281010, 2017AD22006, 2018GXNSFGA281007).

\clearpage
\begin{figure}
\centering
\includegraphics[width=0.8\textwidth]{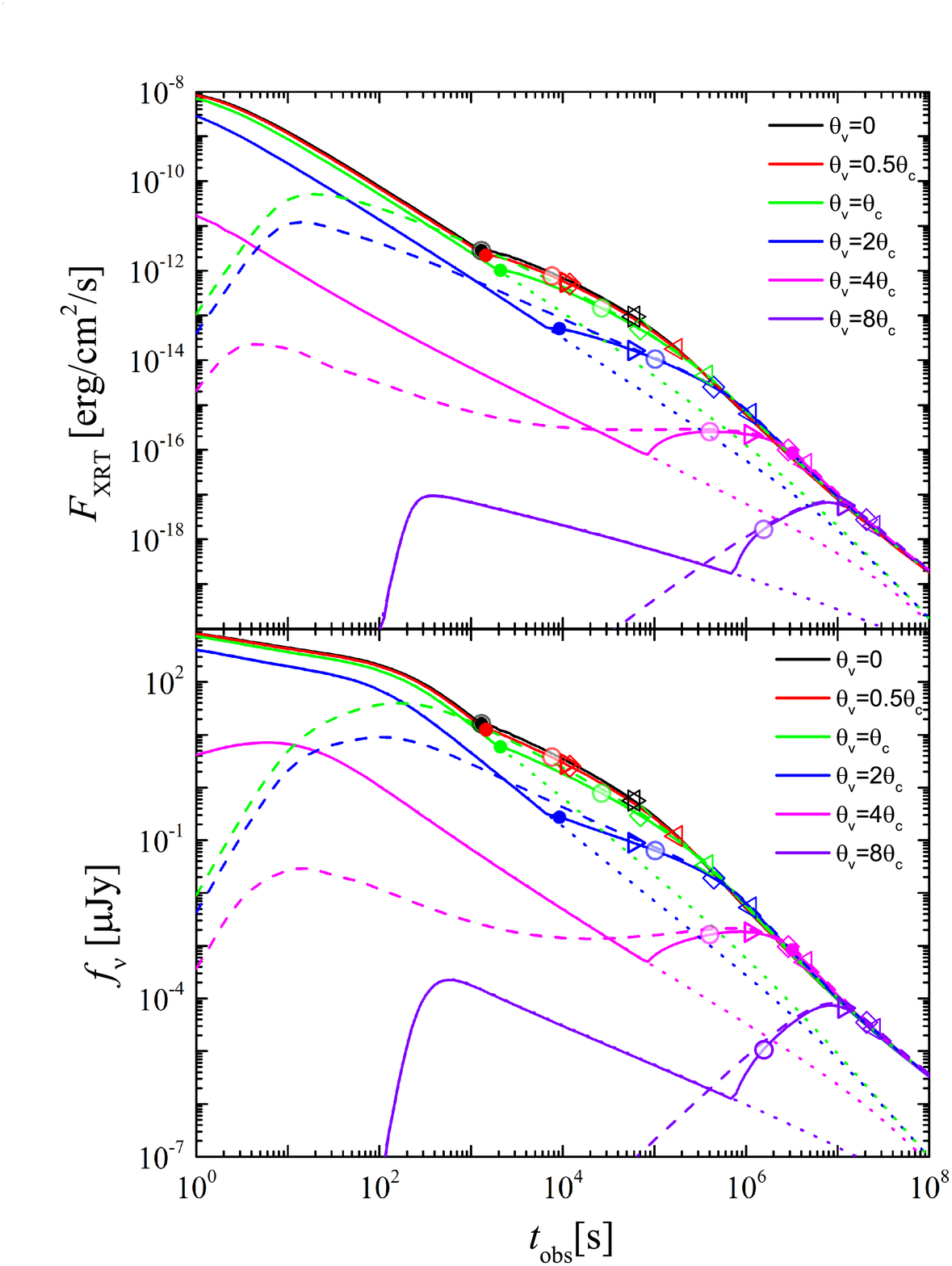}
\caption{
X-ray (top panel, solid lines) and optical (bottom panel, solid lines) light-curves of the external-forward shock
in a free-to-shocked wind environment for different viewing angle $\theta_{\rm v}$,
where $E_{\rm k,iso,on}=10^{52}$~erg, $\Gamma_0=250$,
$\theta_{\rm c}=5^\circ$, $\theta_{\rm jet}=8\theta_{\rm c}$,
$p=2.2$, $\epsilon_{e}=0.1$, $\epsilon_{B}=10^{-4}$, $A_{*}=0.01$, $R_{\rm {tr}}=10^{17} \rm {cm}$, and $z=1$
are adopted.
The dotted/dashed lines represent the afterglows in the pure free/shocked-wind environment
with the same parameters as those adopted to calculating the solid lines.
The symbols of ``{\Large $\bullet$}'' and ``{\LARGE $\circ$}'' represent
the observed time for ($\theta_{\rm v}$, $\varphi_{\rm v}$)-jet and ($0$, $0$)-jet crossing the $R_{\rm tr}$, respectively.
The symbols of ``{\Large $\triangleright$}'', ``{\Large $\diamond$}'', and ``{\Large $\triangleleft$}''
represent the observed time of ($\theta_{\rm c}$, $\varphi_{\rm v}$)-jet,
($0$, $0$)-jet, and ($\theta_{\rm c}$, $\varphi_{\rm v}+\pi$)-jet
beginning to be visible for the observer, respectively.
Here, the observed time for ($\theta_{\rm v}$, $\varphi_{\rm v}$)-jet
crossing the $R_{\rm tr}$ for $\theta_{\rm v}=8\theta_{\rm c}$ is significantly larger than $10^8$~s and is not plotted in the figure.
One can find that the late simultaneous plateaux/bumps appear in the free-to-shocked-wind afterglows.
In addition, these late simultaneous bumps/plateaux are directly followed by a post-jet-break-phase and preceded by a shallow decay.
}\label{MyFigA}
\end{figure}

\begin{figure}
\centering
\begin{tabular}{c}
\includegraphics[width=0.8\textwidth]{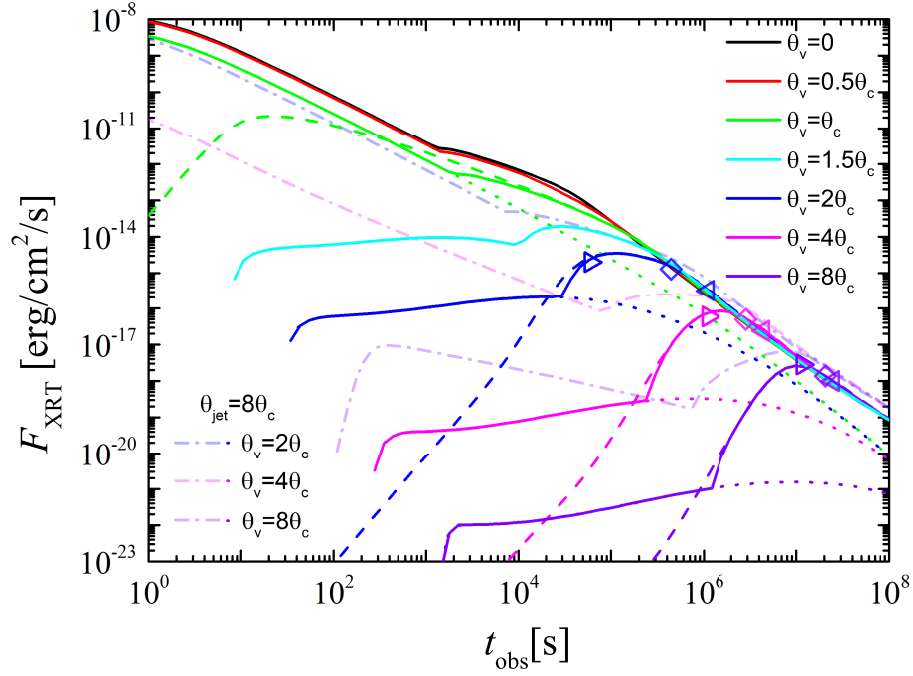}
\end{tabular}
\caption{ X-ray light-curves (solid/dashed/dotted lines) of the external-forward shock
in a free-to-shocked/free/shocked wind environment for different viewing angle $\theta_{\rm v}$,
where the parameters adopted to estimate the emission of the external-forward shock are the same as those in
Figure~{\MyFigA} but with $\theta_{\rm jet}=\theta_{\rm c}$,
and the meanings of ``{\Large $\triangleright$}'', ``{\Large $\diamond$}'', and ``{\Large $\triangleleft$}'' symbols are the same as those in Figure~{\MyFigA}.
For comparison, the solid lines in the top panel of Figure~{\MyFigA} for $\theta_{\rm v}=2\theta_{\rm c}$,
$4\theta_{\rm c}$, and $8\theta_{\rm c}$ are also plotted
with dot-dashed lines in this figure.
One can find that the opening angle of the jet may affect the light-curves before the late simultaneous bumps/plateaux.
}\label{MyFigB}
\end{figure}

\begin{figure}
\centering
\includegraphics[width=0.8\textwidth]{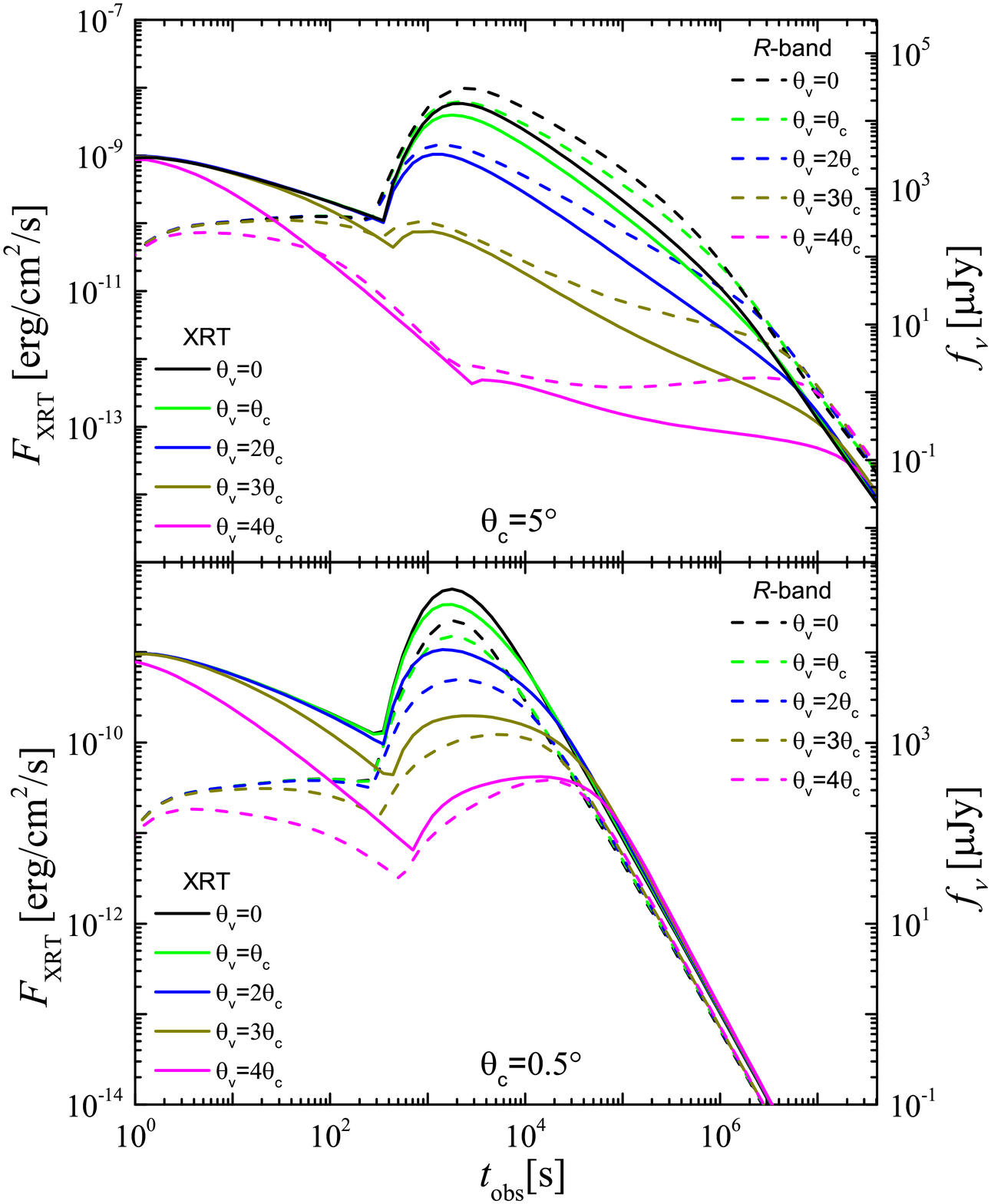}
\caption{
X-ray (solid lines) and optical (dashed lines) light-curves of the external-forward shock
in a free-to-shocked wind environment for different viewing angle $\theta_{\rm v}$,
where a structured jet with $E_{\rm k,iso,on}=10^{55}$erg, $\Gamma_0=100$,
$\theta_{\rm jet}=8\theta_{\rm c}$,
$p=2.2$, $\epsilon_{e}=0.1$, $\epsilon_{B}=10^{-4}$, $A_{*}=0.01$, $R_{\rm {tr}}=10^{17} \rm {cm}$,
and $\theta_{\rm c}=5^\circ$ (top panel) or $\theta_{\rm c}=0.5^\circ$ (bottom panel) is adopted.
For the given structured jet,
the deceleration radius of the on-axis jet flow is beyond the free-wind boundary.
It is found that late broad bumps preceded by a long plateau or shallow decay
indeed appear in the afterglows for the cases with low viewing angle.
In addition, the late broad bump is followed by a normal/steep decay
if a structured jet with high/low characteristic angle $\theta_{\rm c}$
(e.g., $\theta_{\rm c}=5^\circ$/$\theta_{\rm c}=0.5^\circ$) is adopted.
}\label{MyFigC}
\end{figure}

\begin{figure}
\centering
\begin{tabular}{c}
\includegraphics[angle=0,width=0.8\textwidth]{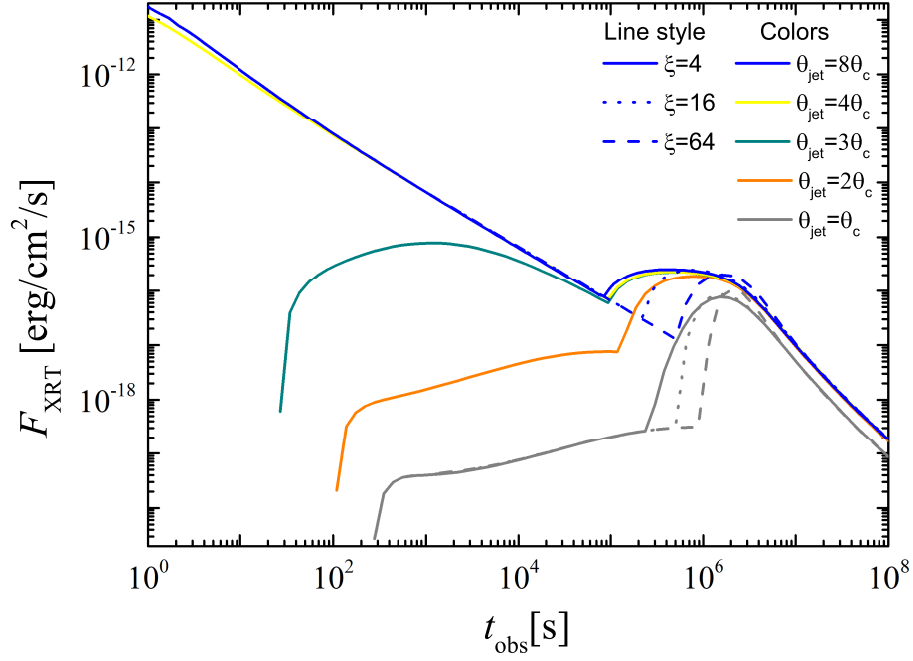}
\end{tabular}
\caption{ X-ray light-curves (solid lines) of the external-forward shock
in a free-to-shocked wind environment for the situations with different $\theta_{\rm jet}$,
where $\theta_{\rm jet}=\theta_{\rm c}$, $2\theta_{\rm c}$, $3\theta_{\rm c}$,
$4\theta_{\rm c}$, and $8\theta_{\rm c}$ are adopted.
The other parameters adopted to estimate the emission of the external-forward shock are
the same as those in Figure~{\MyFigA} and $\theta_{\rm v}=4\theta_{\rm c}$ is set.
It reveals that the early phase of the afterglow is shaped by the emission from the jet flow closing to the line of sight and the post-jet-break phase is shaped by the emission from the jet core.
To test the effect of the density jump factor $\xi$ on the light-curves,
the X-ray afterglows in the situations with $\xi=16$ (dotted lines) and $\xi=64$ (dashed lines)
for $\theta_{\rm jet}=8\theta_{\rm c}$ or $\theta_{\rm jet}=\theta_{\rm c}$ are also shown.
It can be found that the transition from the free-wind-phase to the shocked-wind-phase appears later
and thus the rise of the bump becomes steeper by increasing the value of $\xi$.
}\label{MyFigD}
\end{figure}

\clearpage
\begin{figure}
\centering
\includegraphics[angle=0,width=0.8\textwidth]{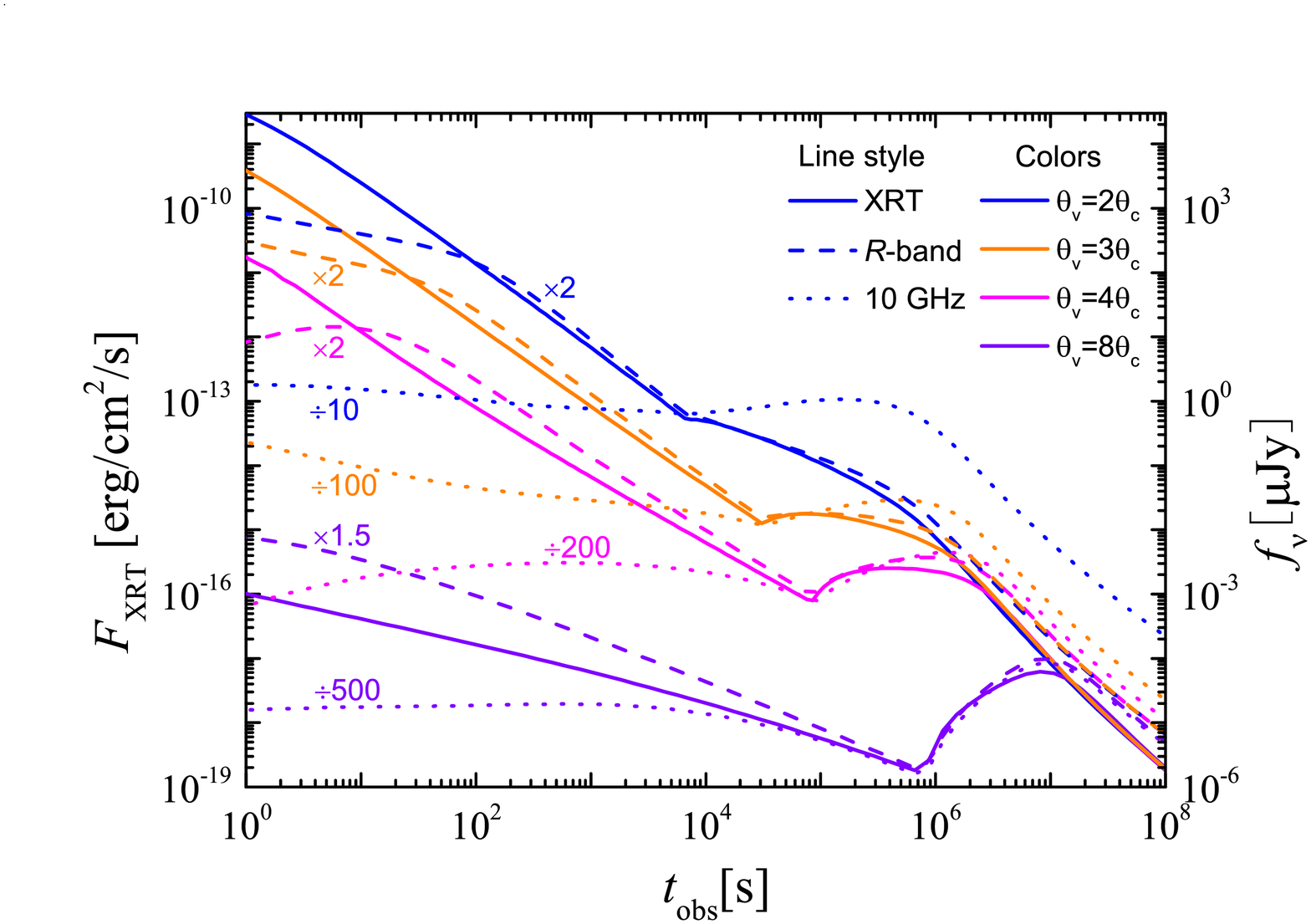}
\caption{
XRT/Optical/Radio afterglows (solid/dashed/dotted lines) from the external-forward shock
in a free-to-shocked wind environment for the situations with
$\theta_{\rm v}=2\theta_{\rm v}$ (blue lines), $3\theta_{\rm v}$ (orange lines), $4\theta_{\rm v}$ (magenta lines)
and $8\theta_{\rm v}$ (violet lines), where the other parameters adopted to estimate the emission of the external-forward shock are the same as those in Figure~{\MyFigA}.
For comparison, the optical/radio afterglow has been shifted by timing a certain factor.
It reveals that the bumps/plateaux in different bands are simultaneous but may be not absolutely achromatic.
}\label{MyFigE}
\end{figure}

\begin{figure}
\centering
\includegraphics[width=0.8\textwidth]{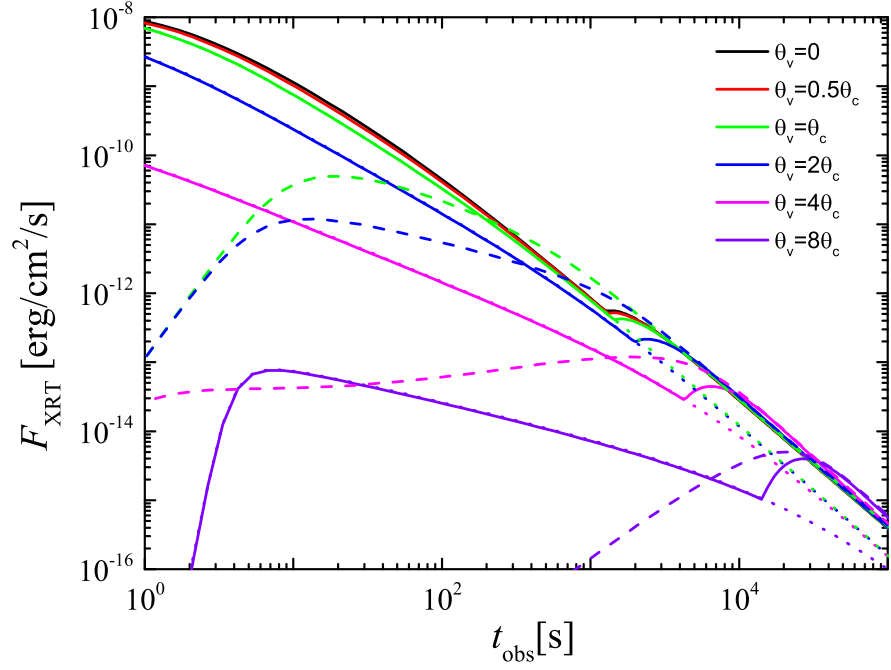}
\caption{Same as Figure~{\MyFigA} but with $\theta_{\rm c}=0.5^\circ$.
 The width of late bump/plateau which
is narrower than that in Figure~{\MyFigA},
which reveals that the characteristic angle $\theta_{\rm c}$ of the structured jet
 may affect the width of the late bump/plateau for a same $\theta_{\rm v}/\theta_{\rm c}$.
}\label{MyFigF}
\end{figure}

\clearpage
\begin{figure}
\centering
\begin{tabular}{c}
\includegraphics[trim=0 30 0 40, angle=0,width=0.65\textwidth]{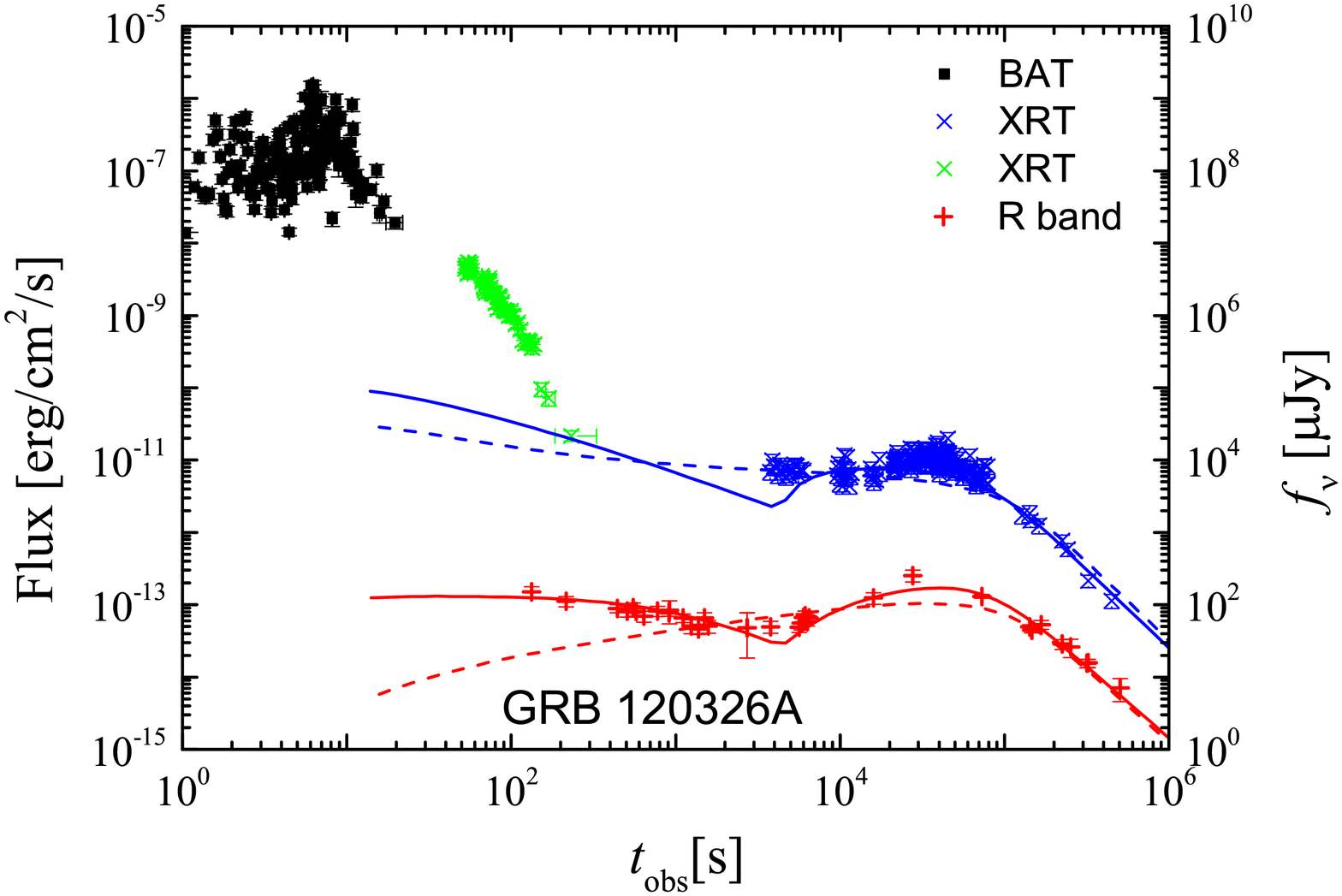}\\
\includegraphics[trim=0 30 0 30, angle=0,width=0.65\textwidth]{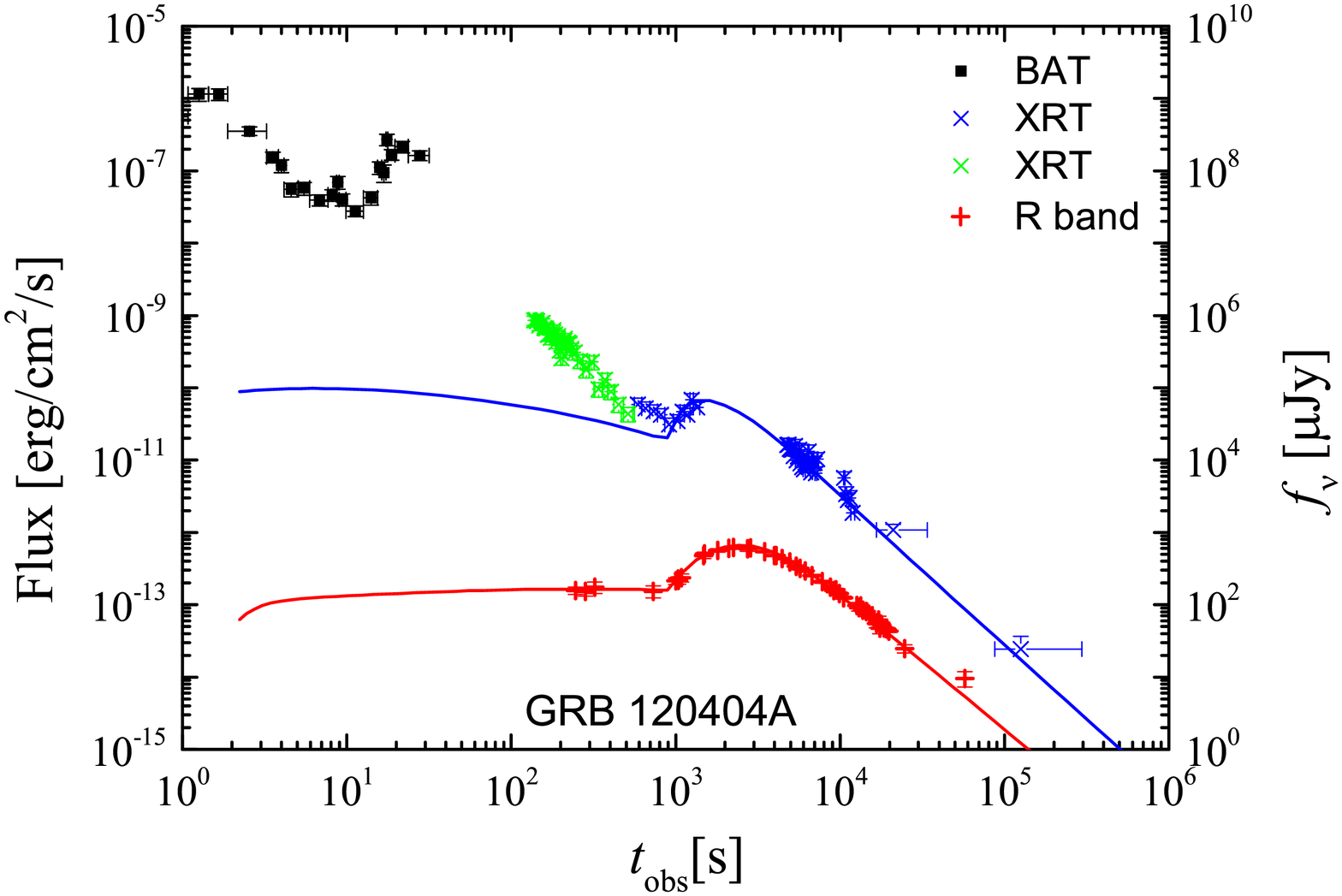}\\
\includegraphics[trim=0 30 0 30, angle=0,width=0.65\textwidth]{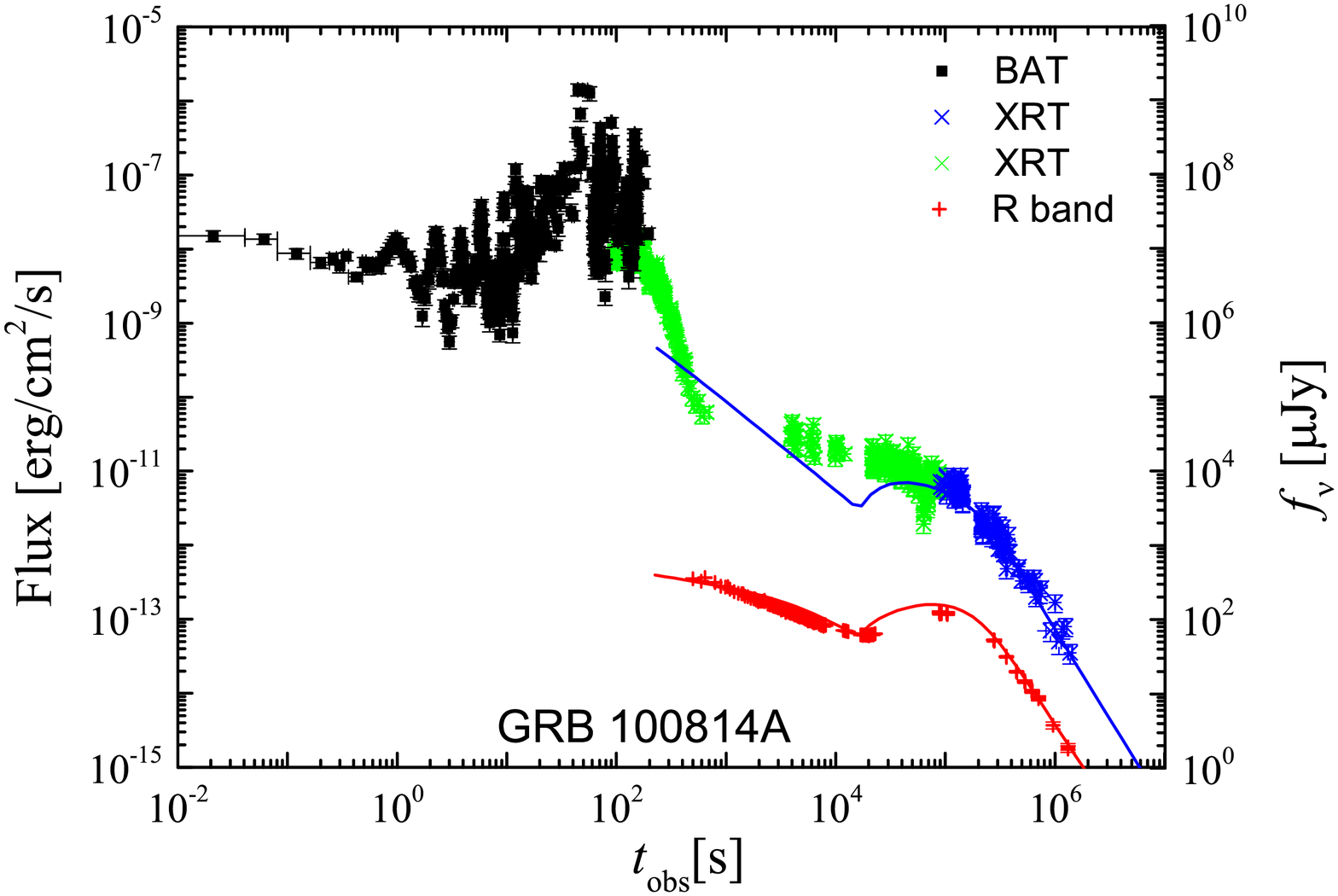}\\
\end{tabular}
\caption{Fitting result of the afterglows in GRB~120326A (top panel), GRB~120404A (middle panel), and GRB~100814A (bottom panel), where blue ``$\times$'' and red ``$+$'' symbols represent the XRT and optical data used in our fittings, respectively. The blue and red lines are the optimal fitting of the late XRT and optical afterglows, respectively.
}\label{MyFigG}
\end{figure}
%

\clearpage
\begin{figure}
\centering
\begin{tabular}{c}
\includegraphics[width=\textwidth]{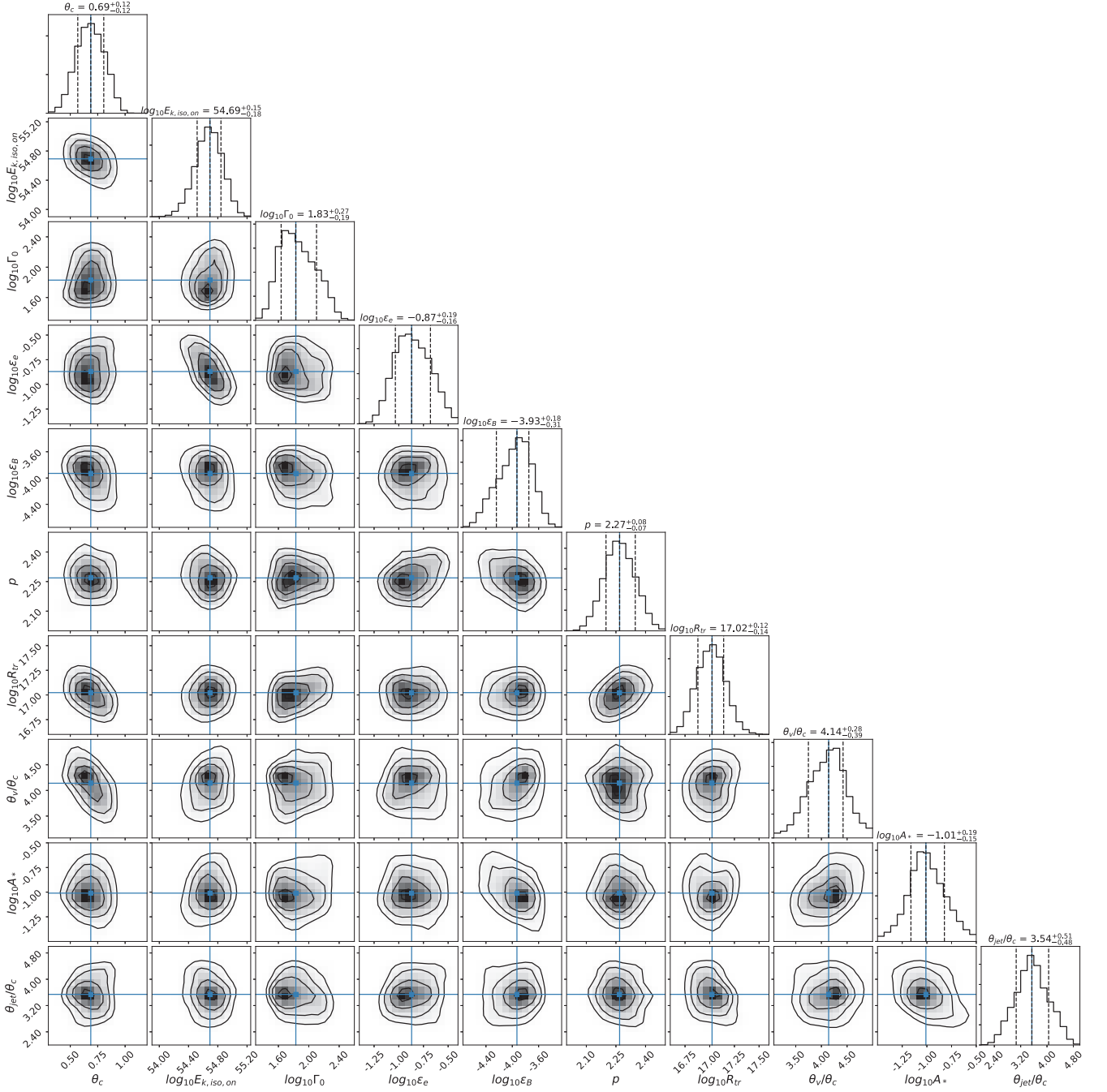}
\end{tabular}
\caption{Posterior probability density functions for the physical parameters of the external-forward shock in GRB 120326A from MCMC simulations.
}\label{MyFigH}
\end{figure}

\clearpage
\begin{table}
\caption{Parameters Estimated from the MCMC Sampling}\label{MyTabA}
\centering{
\begin{tabular}{cccc}
\hline \hline
Parameters & GRB~120326A & GRB~120404A & GRB~100814A\\
\hline
$\theta_{\rm c}$ & $0.69^{+0.12}_{-0.12}$   & $0.58^{+0.14}_{-0.11}$ & $0.89^{+0.12}_{-0.05}$   \\

$\log_{10}(E_{\rm k,iso, on}/\rm erg)$ & $54.69^{+0.15}_{-0.18}$ & $53.78^{+0.16}_{-0.14}$ & $54.89^{+0.07}_{-0.13}$\\

${\log_{10}}\Gamma_0$ & $1.83^{+0.27}_{-0.19}$   & $1.86^{+0.03}_{-0.04}$ & $2.28^{+0.11}_{-0.09}$\\

${\log_{10}}\epsilon_{e}$ & $-0.87^{+0.19}_{-0.16}$  & $-1.01^{+0.12}_{-0.09}$ & $-0.51^{+0.01}_{-0.10}$\\

${\log_{10}}\epsilon_{B}$ & $-3.93^{+0.18}_{-0.31}$  & $-3.21^{+0.16}_{-0.18}$ & $-4.69^{+0.17}_{-0.10}$\\

$p$ & $2.27^{+0.08}_{-0.07}$ & $2.29^{+0.06}_{-0.07}$ & $2.60^{+0.04}_{-0.05}$ \\

$\log_{10}(R_{\rm tr}/{\rm cm})$ & $17.02^{+0.12}_{-0.14}$  & $16.80^{+0.06}_{-0.08}$ & $17.60^{+0.07}_{-0.13}$\\

$\theta_{\rm v}/\theta_{\rm c}$ & $4.14^{+0.28}_{-0.39}$  & $0.68^{+0.28}_{-0.25}$ &$3.66^{+0.09}_{-0.05}$\\

${\log_{10}}A_{*}$ & $-1.01^{+0.19}_{-0.15}$  & $-1.46^{+0.11}_{-0.11}$ &$-0.66^{+0.11}_{-0.03}$\\

$\theta_{\rm {jet}}/\theta_{\rm c}$ & $3.54^{+0.51}_{-0.48}$  &  $1.17^{+0.12}_{-0.10}$ & $4.03^{+0.90}_{-0.44}$ \\
\hline
$E_{\rm k}/\rm erg$ & $2.23\times 10^{51}$  &  $9.63\times 10^{49}$ & $5.88\times 10^{51}$ \\
\hline
\end{tabular}\\}
\tablenotemark{}{ $E_{\rm k}$ is the total kinetic energy of the structured jet.}
\end{table}

\clearpage
{
\appendix\label{Sec:Appendix}
\section{Analytical Solutions of the External-forward Shock and its Emission}
\subsection{Analytical Solution for the Dynamics of the External-forward Shock}
The emission of the external-forward shock is related to
$\gamma'_{\rm e,m}$, $\gamma'_{\rm e,c}$, $B'$, $\Gamma$, and $\rho$ (or $R$).
In addition,
the equations of $\gamma'_{\rm e,m} =\epsilon _e(p - 2){m_{\rm{p}}}\Gamma/[(p - 1){m_{\rm{e}}}]$,
$\gamma'_ {\rm e,c}=6\pi m_{\rm e} c/(\sigma_{\rm T}\Gamma {B'}^2 t_{\rm obs}^{\rm on})$,
and $B'=[32\pi \rho(R) \epsilon_B]^{1/2} \Gamma c$
are used to estimate $\gamma'_{\rm e,m}$, $\gamma'_{\rm e,c}$, and $B'$, respectively.
Then, the emission of the external-forward shock at observer time $t_{\rm obs}$ can be estimated
only if the dependence of $R=R_{\rm obs}(\theta, \varphi, \theta_{\rm v})$
and $\Gamma_{\rm obs}(\theta, \varphi, \theta_{\rm v})=\Gamma(\theta, R_{\rm obs})$ on $t_{\rm obs}$ are presented,
where $R_{\rm obs}$ and $\Gamma_{\rm obs}$ are the location and Lorentz factor of the external-forward shock
for the ($\theta$, $\varphi$)-jet at observer time $t_{\rm obs}$.
The $t_{\rm obs}$-dependent $\Gamma_{\rm obs}$ and $R_{\rm obs}$ can be described as
\begin{eqnarray}\label{Eq:AP_Dynamics_On-axis}
\left\{ \begin{array}{l}
\displaystyle \frac{\Gamma_{\rm obs}}{{{\Gamma _0}}} = {\left[ {1 + {{\left( {\zeta \frac{{{t_{{\rm{obs}}}}}}{{{t_{{\rm{dec}}}}}}} \right)}^{3 - s}}} \right]^{\frac{1}{{ \epsilon  + 2s - 8}}}},\\
\displaystyle \frac{R_{\rm obs}}{{{R_{{\rm{dec}}}}}} = {\left[ {1 + {\zeta ^{\frac{{ \epsilon  - 2}}{{6 - 2s}}}}\frac{{{t_{{\rm{obs}}}}}}{{{t_{{\rm{dec}}}}}}} \right]^{\frac{{6 - 2s}}{{ \epsilon  + 2s - 8}}}}\frac{{{t_{{\rm{obs}}}}}}{{{t_{{\rm{dec}}}}}},\\
\displaystyle D = 2{\Gamma _0}\left( {\frac{\Gamma }{{{\Gamma _0}}}} \right),
\end{array} \right.\;\;\;\;\;\;{\rm{for}}\;\;2\Gamma\sqrt \zeta\sin (\Theta /2) < 1,
\end{eqnarray}
and
\begin{eqnarray}\label{Eq:AP_Dynamics_Off-axis}
\left\{ \begin{array}{l}
\displaystyle \frac{\Gamma_{\rm obs}}{{{\Gamma _0}}} = {\left[ {1 + {{\left( {a\frac{{{t_{{\rm{obs}}}}}}{{{t_{{\rm{dec}}}}}}} \right)}^{3 - s}}} \right]^{\frac{1}{{ \epsilon  - 2}}}},\\
\displaystyle \frac{R_{\rm obs}}{{{R_{{\rm{dec}}}}}} = a\frac{{{t_{{\rm{obs}}}}}}{{{t_{{\rm{dec}}}}}},\\
\displaystyle D = 2{\Gamma _0}a{\left( {\frac{\Gamma }{{{\Gamma _0}}}} \right)^{ - 1}},
\end{array} \right.\;\;\;\;\;\;{\rm{for}}\;\;2\Gamma\sqrt \zeta \sin (\Theta /2) >1,
\end{eqnarray}
where the medium mass density with $\rho=\rho_0 R^{-s}$ is adopted and the other parameters are
\begin{equation}
{t_{{\rm{dec}}}} = (1 + z)\frac{{R_{{\rm{dec}}}}}{{2\Gamma _0^2c}},
\end{equation}
\begin{equation}\label{Eq:R_dec}
{R_{{\rm{dec}}}}={\left[ {\frac{{4\pi \rho_0{\Gamma _0}(2 -  \epsilon )}}{(3 - s)M_0(\theta)}} \right]^{1/(s - 3)}},
\end{equation}
\begin{equation}
\zeta=(\epsilon  - 8 + 2s)/(\epsilon  - 2),
\end{equation}
\begin{equation}
a = 1/[(1 - \cos \Theta )2\Gamma _0^2],
\end{equation}
\begin{equation}\label{Eq:t_obson}
t_{\rm obs}^{{\rm{on}}} \approx \frac{R_{\rm obs}}{{2\Gamma _0^2c{{\left[ {1 + {\zeta ^{(\varepsilon {\rm{ - 2}}){\rm{/2}}}}{{\left( {R_{\rm obs}/{{{R_{{\rm{dec}}}}}}} \right)}^{3 - s}}} \right]}^{\frac{2}{{\varepsilon  - 2}}}}}}.
\end{equation}
The derivation of Equations~(\ref{Eq:AP_Dynamics_On-axis}), (\ref{Eq:AP_Dynamics_Off-axis}) and
(\ref{Eq:t_obson}) are presented as follows.

For the ($\theta$, $\varphi$)-jet, the evolution of the bulk Lorentz factor
in the relativistic regime can be estimated with
\begin{equation}\label{Eq:AP_dynamic}
{\frac{m(R)}{M_0(\theta)}=-(\Gamma_0-1)^{1/2}(\Gamma_0+1)^{1/{2-\epsilon}}\int_{\Gamma_0}^{\Gamma(\theta, R)}
(\gamma-1)^{-3/2}(\gamma+1)^{-3/2+\epsilon}d\gamma
},
\end{equation}
where $\Gamma_0$ is the initial Lorentz factor of the fireball.
Equation~(\ref{Eq:AP_dynamic}) is obtained based on the following equations (\citealp{Chiang_J-1999-Dermer_CD-ApJ.512.699C,Piran_T-1999-PhR.314.575P}):
\begin{eqnarray}\label{Eq:Lorentz_Evo}
\frac{{d{\Gamma }}}{{d{m}}} =  - \frac{{\Gamma ^2 - 1}}{{M'}},
\end{eqnarray}
$M'=M_0(\theta)+m(R)+U'/c^2$, and
$d{U'}=(1-\epsilon)(\Gamma-1)c^2dm$.
Since Equation~(\ref{Eq:Lorentz_Evo}) fails to reproduce the dynamics in the
non-relativistic regime,
\cite{Huang_YF-1999-Dai_ZG-MNRAS.309.513H,Huang_YF-2000-Gou_LJ-ApJ.543.90H}
revised Equation~(\ref{Eq:Lorentz_Evo}) and proposed Equation~(\ref{eq:Gamma})
to more correctly delineate the blastwave dynamics.
Then, Equation~(\ref{Eq:Lorentz_Evo}) is only applied for the relativistic regime.
In the relativistic regime, i.e., $\Gamma\gg 1$, Equation~(\ref{Eq:AP_dynamic}) is reduced to
\begin{equation}\label{Eq:dynamic}
\frac{{{\Gamma }}}{{{\Gamma _0}}} = {\left[ {1 + (2 -  \epsilon )\frac{{{\Gamma _0}{m}}}{{{M_0}}}} \right]^{\frac{1}{{ \epsilon  - 2}}}}.
\end{equation}
With ${m} = \int_{{R_0}}^{R}{{r^2}\rho dr}  \approx \rho_0 {R^{3 - s}}/(3 - s)$, one can have
\begin{equation}\label{Eq:AP_Gamma}
\frac{{{\Gamma }}}{{{\Gamma _0}}} = {\left[ {1 + {{\left( {\frac{R}{{{R_{{\rm{dec}}}}}}} \right)}^{3 - s}}} \right]^{\frac{1}{{ \epsilon  - 2}}}}.
\end{equation}
Based on Equation~(\ref{Eq:AP_Gamma}) and $t_{{\rm obs}}^{\rm on}(\theta, R)=\int_{R_0}^{R} {(c - \upsilon)} {{dr}}/{c\upsilon}$, we can have
\begin{equation}\label{Eq:t_obs_on1}
t_{{\rm obs}}^{\rm on} \approx \left\{ {\begin{array}{*{20}{c}}
\displaystyle{\frac{R}{{2\Gamma _0^2c}},}&{R < {R_{{\rm{dec}}}},}\\
\displaystyle {\frac{{\varepsilon  - 2}}{{\varepsilon  - 8 + 2s}}\frac{R}{{2{\Gamma ^2}c}},}&{R \gg {R_{{\rm{dec}}}},}
\end{array}} \right.
\end{equation}
and approximately take
\begin{equation}
t_{\rm{obs}}^{{\rm{on}}} \approx \frac{R}{{2\Gamma _0^2c{{\left[ {1 + {\zeta ^{(\varepsilon {\rm{ - 2}}){\rm{/2}}}}{{\left( {R/{{{R_{{\rm{dec}}}}}}} \right)}^{3 - s}}} \right]}^{\frac{2}{{\varepsilon  - 2}}}}}}.
\end{equation}
Correspondingly, we can approximately take
\begin{equation}\label{Eq:AP_R}
\frac{R}{{{R_{{\rm{dec}}}}}} = {\left[ {1 + {\zeta ^{\frac{{ \epsilon  - 2}}{{6 - 2s}}}}\frac{{t_{{\rm obs}}^{{\rm{on}}}}}{{{t_{{\rm{dec}}}}}}} \right]^{\frac{{6 - 2s}}{{ \epsilon  + 2s - 8}}}}\frac{{t_{{\rm obs}}^{{\rm{on}}}}}{{{t_{{\rm{dec}}}}}},
\end{equation}
where ${t_{{\rm{dec}}}}$ is the deceleration time of the ($\theta$, $\varphi$)-jet
while it moves in the line of sight.
With Equation~(\ref{Eq:AP_R}), Equation~(\ref{Eq:AP_Gamma}) can be approximately described as
\begin{equation}\label{Eq:AP_Lorentz_Factor}
\frac{\Gamma }{{{\Gamma _0}}} = {\left[ {1 + {{\left( {\zeta \frac{{t_{\rm obs}^{\rm on}}}{{{t_{{\rm{dec}}}}}}} \right)}^{3 - s}}} \right]^{\frac{1}{{ \epsilon  + 2s - 8}}}}.
\end{equation}

In addition, Equations~(\ref{Eq:tobs}) and (\ref{Eq:t_obs_on1}) reveal
\begin{equation}\label{Eq:AP_tobs}
{t_{{\rm{obs}}}}(\theta, \varphi, \theta_{\rm v}, R)\approx \left\{ {\begin{array}{*{20}{c}}
{(1 - \cos \Theta )R(1 + z)/c,\;\;\;\;}&{2\Gamma\sqrt \zeta\sin (\Theta /2) >1,}\\
{t_{{\rm{obs}}}^{{\rm{on}}},\;\;\;\;}&{2\Gamma\sqrt \zeta\sin (\Theta /2) <1.}
\end{array}} \right.
\end{equation}
For $2\Gamma \sqrt \zeta\sin (\Theta /2) < 1$,
Equation~(\ref{Eq:AP_Dynamics_On-axis}) can be easy derived
based on Equations~(\ref{Eq:AP_R}), (\ref{Eq:AP_Lorentz_Factor}), and (\ref{Eq:AP_tobs}).
For $2\Gamma \sqrt \zeta\sin (\Theta /2) > 1$,
one can have ${R_{\rm obs}}/{{{R_{{\rm{dec}}}}}} = a{{{t_{{\rm{obs}}}}}}/{{{t_{{\rm{dec}}}}}}$
and thus the $t_{\rm obs}$-dependent $\Gamma$ can be derived based on Equation~(\ref{Eq:AP_Gamma}).

\subsection{$P'(\nu',\theta,R)$ for the ($\theta$, $\varphi$)-jet}
Assuming the electron energy spectrum as $n'_{\rm e}(\gamma'_{\rm e}, \theta, R)
= {N_{{\rm{e}},{\rm{tot}}}}{(1 - k)( {\gamma'_{\rm e}/{\gamma'_{\min }}})^{ - k}}/(4\pi{\gamma '_{\min }})$ at ${\gamma'_e} \gtrsim {\gamma '_{{\rm{min}}}} \equiv \min ({\gamma' _{\rm{e,c}}},{\gamma'_{\rm{e,m}}})$, one can have
\begin{eqnarray}
P'(\nu',\theta, R)={\left( {\frac{{\nu '}}{{\nu'_{\min}}}} \right)^{1/3}}{{F_{\nu ,\max }'}}\;\;{\rm for}\;\;
\nu ' < {\nu'_{\min }}
\end{eqnarray}
without considering the synchronization self-absorption effect,
where $\nu'_{\min} ={\gamma'_{\min}}^2{q_{\rm e}}B'/(2\pi {m_{\rm{e}}}c)$,
${{F_{\nu ,\max }'}}=\xi {P_{\nu ,\max }}{N_{e,{\rm{tot}}}}/(4\pi\Gamma)$
with $\xi =2.15\times (3k - 3)/(3k - 1)\sim 1.5$ for $2\lesssim k \lesssim 4$,
${P_{\nu ,\max }}(\Gamma_{\rm obs}, R_{\rm obs}) = {m_{\rm e}}{c^2}{\sigma _{\rm T}}B'\Gamma/(3{q_{\rm{e}}})$,
and ${N_{{\rm{e}},{\rm{tot}}}}(R_{\rm obs}) = 4\pi m/{m_{\rm{p}}}$.

The exact electron spectrum depend on the injection rate $Q'$, i.e., Equation(\ref{Eq:Q}),
of electrons from the shock and the cooling effect.
Considering only the synchrotron radiation cooling effect,
the steady state of electron spectrum can be described as
\[
n'_{\rm e}(\gamma'_{\rm e},\theta,R) \propto
\left\{ {\begin{array}{*{20}{c}}
\gamma_{\rm e}^{\prime {-2}}, &{{{\gamma}_{\rm e,c}^\prime} < {{\gamma}_{\rm e}^\prime} < {{\gamma_{\rm e,m}^\prime}}},\\
{\gamma_{\rm e}^{\prime { - (p + 1)}},}&{{{\gamma}_{\rm e}^\prime} > {{\gamma}_{\rm e,m}'}},
\end{array}} \right.
\]
for fast cooling (i.e, $\gamma'_{\min}=\gamma'_{\rm e,c}<\gamma'_{\rm e,m}$),
and
\[
n'_{\rm e}(\gamma'_{\rm e},\theta,R) \propto
\left\{ {\begin{array}{*{20}{c}}
\gamma_{\rm e}^{\prime {-p}}, &{{{\gamma}_{\rm e,m}^\prime} < {{\gamma}_{\rm e}'}< {{\gamma}_{\rm e,c}^\prime}},\\
{\gamma_{\rm e}^{\prime { - (p + 1)}},}&{{{\gamma}_{\rm e}'} > {{\gamma}_{\rm e,c}^\prime}},
\end{array}} \right.
\]
for slow cooling (i.e, $\gamma'_{\min}=\gamma'_{\rm e,m}<\gamma'_{\rm e,c}$).
Correspondingly, one can have
\begin{eqnarray}\label{Eq:FastCooling_CoM}
P'(\nu',\theta,R)=\left\{ {\begin{array}{*{20}{c}}
{{{\left( {{\nu'}/{{{{\nu_{\rm c}'}}}}} \right)}^{1/3}}F_{\nu ,\max }',}&{\nu ' < {{\nu_{\rm c}'}}}\\
{{{\left( {{{\nu '}}/{{{{\nu_{\rm c}'}}}}} \right)}^{ - 1/2}}F_{\nu ,\max }',}&{{{\nu_{\rm c}'}} < \nu ' < {{\nu_{\rm m}'}}}\\
{{{\left( {\nu_{\rm m}'/{{{{\nu_{\rm c}'}}}}} \right)}^{ - 1/2}}{{\left( {{{\nu '}}/{{{{\nu_{\rm m}'}}}}} \right)}^{ - p/2}}F_{\nu ,\max }',}&{\nu ' > {{\nu_{\rm m} '}}}
\end{array}} \right.
\end{eqnarray}
for fast cooling, and
\begin{eqnarray}\label{Eq:SlowCooling_CoM}
P'(\nu',\theta,R)=\left\{ {\begin{array}{*{20}{c}}
{{{\left( {{{\nu '}}/{{{{\nu_{\rm m}'}}}}} \right)}^{1/3}}F_{\nu ,\max }',}&{\nu ' < {{\nu_{\rm m}'}}}\\
{{{\left( {{{\nu '}}/{{{{\nu_{\rm m}'}}}}} \right)}^{ - (p - 1)/2}}F_{\nu ,\max }',}&{{{\nu_{\rm m}'}} < \nu ' < {{\nu_{\rm c}'}}}\\
{{{\left( {{{{\nu_{\rm c}'}}}/{{{{\nu_{\rm m}'}}}}} \right)}^{ - (p - 1)/2}}{{\left( {{{\nu '}}/{{{{\nu_{\rm c}'}}}}} \right)}^{ - p/2}}F_{\nu ,\max }',}&{\nu ' > {{\nu_{\rm c}'}}}
\end{array}} \right.
\end{eqnarray}
for slow cooling,
where $\nu'_{\rm c} ={\gamma'_{\rm c}}^2{q_{\rm e}}B'/(2\pi {m_{\rm{e}}}c)$
and $\nu'_{\rm m} ={\gamma'_{\rm m}}^2{q_{\rm e}}B'/(2\pi {m_{\rm{e}}}c)$.

Based on Equations~(\ref{Eq:FastCooling_CoM}), (\ref{Eq:SlowCooling_CoM}),
(\ref{EQ:ObsFlux}), and (\ref{EQ:ObsFluxTotal}),
one can estimate the observed total flux density $f_{\nu} ({t_{{\rm{obs}}}})$ from the external-forward shock.

\subsection{Analytical Light-curves for A Top-hat Jet}\label{Sec_LC_TopHat}
In order to estimate the analytical light-curves for a top-hat jet,
we assume the radiation energy of the jet shell at radius $R$ are all observed at the same time.
For a top-hat jet, the observed flux from the jet shell is mainly dominated by
that from the jet flow being close to the line of sight.
Then, the location and Lorentz factor of the jet flow being closest to the line of sight at the observer time $t_{\rm obs}$ can be used to depict the emission of the external-forward shock
and are represented with $\bar{R}$ and $\bar{\Gamma}$, respectively.
With Equations~(\ref{Eq:FastCooling_CoM}) and (\ref{Eq:SlowCooling_CoM}), one can have
\begin{eqnarray}\label{Eq:FastCooling_OnAxis}
{f_\nu }({t_{{\rm{obs}}}})
=
\left\{ {\begin{array}{*{20}{c}}
{{{\left( {\nu /{\nu _{\rm{c}}}} \right)}^{1/3}}F_{\nu ,\max },}&{\nu  < {\nu _{\rm{c}}},}\\
{{{\left( {\nu /{\nu _{\rm{c}}}} \right)}^{ - 1/2}}{F_{\nu ,\max }},}&{{\nu _{\rm{c}}} < \nu  < {\nu _{\rm{m}}},}\\
{{{\left( {{\nu _{\rm{m}}}/{\nu _{\rm{c}}}} \right)}^{ - 1/2}}{{\left( {\nu /{\nu _{\rm{m}}}} \right)}^{ - p/2}}{F_{\nu ,\max }},}&{\nu  > {\nu _{\rm{m}}},}
\end{array}} \right.
\end{eqnarray}
for the fast cooling case, and
\begin{eqnarray}\label{Eq:SlowCooling_OnAxis}
{f_\nu }({t_{{\rm{obs}}}}) = \left\{ {\begin{array}{*{20}{c}}
{{{\left( {\nu /{\nu _{\rm{m}}}} \right)}^{1/3}}{F_{\nu ,\max }},}&{\nu  < {\nu _{\rm{m}}},}\\
{{{\left( {\nu /{\nu _{\rm{m}}}} \right)}^{ - (p - 1)/2}}{F_{\nu ,\max }},}&{{\nu _{\rm{m}}} < \nu  < {\nu _{\rm{c}}},}\\
{{{\left( {{\nu _{\rm{c}}}/{\nu _{\rm{m}}}} \right)}^{ - (p - 1)/2}}{{\left( {\nu /{\nu _{\rm{c}}}} \right)}^{ - p/2}}{F_{\nu ,\max }},}&{\nu  > {\nu _{\rm{c}}},}
\end{array}} \right.
\end{eqnarray}
for the slow cooling case, where
\textcircled{\footnotesize{I}}
for the situation with $0\leqslant \theta_{\rm v} \lesssim \theta_{\rm jet}$,
$\bar{R}=R_{\rm obs}(\theta_{\rm v}, \varphi_{\rm v}, \theta_{\rm v})$,
$\bar{\Gamma}=\Gamma_{\rm obs}(\theta_{\rm v}, \varphi_{\rm v}, \theta_{\rm v})$,
${\nu _{\rm{m}}} = {\nu'_{\rm{m}}}2\bar{\Gamma} /(1 + z)$,
${\nu _{\rm{c}}} = {\nu'_{\rm{c}}}2\bar{\Gamma}/(1 + z)$,
and ${F_{\nu ,\max }} = \kappa (1 + z){P_{\nu ,\max }}(\bar{\Gamma}, \bar{R}){N_{{\rm{e}},{\rm{tot}}}}(\bar{R})/(4\pi d_{\rm{L}}^2)$ with
$\kappa=\xi \times \{ {1 - 1/{{\{ 1 + 2{\Bar{\Gamma} ^2}[1 - \cos ({\theta _{{\rm{jet}}}} + {\theta _{\rm{v}}})]\} }^2}} \}$;
\textcircled{\footnotesize{II}}
for the situation with $\theta_{\rm v}>\theta_{\rm jet}+1/\bar{\Gamma}$,
$\bar{R}=R_{\rm obs}(\theta_{\rm v}-\theta_{\rm jet}, \varphi_{\rm v}, \theta_{\rm v})$,
$\bar{\Gamma}=\Gamma_{\rm obs}(\theta_{\rm v}-\theta_{\rm jet}, \varphi_{\rm v}, \theta_{\rm v})$,
${\nu _{\rm{m}}} = {\nu'_{\rm{m}}}/[(1 + z)2\bar{\Gamma} ]$,
${\nu _{\rm{c}}} = {\nu'_{\rm{c}}}/[(1 + z)2\bar{\Gamma}]$,
and ${F_{\nu ,\max }} = \kappa (1 + z){P_{\nu ,\max }}(\bar{\Gamma},\bar{R}){N_{{\rm{e}},{\rm{tot}}}}(\bar{R})/(4\pi d_{\rm{L}}^2{\bar{\Gamma}^4})$
with $\kappa  = \xi {\theta _{{\rm{jet}}}}{(\pi \sin {\theta _{\rm{v}}})^{ - 1}}[{({\theta _{\rm{v}}} - {\theta _{{\rm{jet}}}})^{ - 4}} - {({\theta _{\rm{v}}} + {\theta _{{\rm{jet}}}})^{ - 4}}]$.
The analytical result of \textcircled{\footnotesize{I}} is consistent with that of \cite{Sari_R-1998-Piran_T-ApJ.497L.17S}.

The expression of ${f_\nu }({t_{{\rm{obs}}}})$ is derived based on Equation~(\ref{EQ:ObsFluxTotal}),
where
\begin{eqnarray}
{f_\nu }({t_{{\rm{obs}}}}) \approx \frac{{(1 + z)}}{{2d_{\rm{L}}^2}}P'(\nu \frac{{1 + z}}{{2\bar \Gamma }},{\theta _{\rm{v}}},\bar R)\int_0^{{\theta _{{\rm{jet}}}} + {\theta _{\rm{v}}}} {{D^3}\sin \Theta d\Theta }
\end{eqnarray}
or
\begin{eqnarray}\label{Eq:top-hat-onaxis}
{f_\nu }({t_{{\rm{obs}}}}) \approx \frac{{(1 + z)}}{{d_{\rm{L}}^2}}\bar \Gamma P'(\nu \frac{{1 + z}}{{2\bar \Gamma }},{\theta _{\rm{v}}},\bar R)\left\{ {1 - {{\left[ {\frac{1}{{1 + \left[ {1 - \cos ({\theta _{{\rm{jet}}}} + {\theta _{\rm{v}}})} \right]/(1 - \beta )}}} \right]}^2}} \right\}
\end{eqnarray}
is obtained for an observer with $0\leqslant \theta_{\rm v} \lesssim \theta_{\rm jet}$,
and
\begin{eqnarray}
{f_\nu }({t_{{\rm{obs}}}}) \approx \frac{{(1 + z)}}{{4\pi d_{\rm{L}}^2}}P'(\nu \frac{{1 + z}}{{2\bar \Gamma }},{\theta _{\rm{v}}} - {\theta _{{\mathop{\rm jet}\nolimits} }},\bar R)\left( {\frac{{2R{\theta _{{\rm{jet}}}}}}{{R\sin {\theta _{\rm{v}}}}}} \right)\int_{{\theta _{\rm{v}}} - {\theta _{{\rm{jet}}}}}^{{\theta _{\rm{v}}} + {\theta _{{\rm{jet}}}}} {{D^3}\sin \Theta d\Theta }
\end{eqnarray}
or
\begin{eqnarray}\label{Eq:top-hat-off-axis_2}
{f_\nu }({t_{{\rm{obs}}}}) \approx \frac{{(1 + z)}}{{d_{\rm{L}}^2}}P'(\nu \frac{{1 + z}}{{2\bar \Gamma }},{\theta _{\rm{v}}} - {\theta _{{\mathop{\rm jet}\nolimits} }},\bar R)\frac{1}{{{{\bar \Gamma }^3}}}\left( {\frac{{{\theta _{{\rm{jet}}}}}}{{\pi \sin {\theta _{\rm{v}}}}}} \right)\left[ {\frac{1}{{{{({\theta _{{\rm{jet}}}} - {\theta _{\rm{v}}})}^4}}} - \frac{1}{{{{({\theta _{{\rm{jet}}}} + {\theta _{\rm{v}}})}^4}}}} \right]
\end{eqnarray}
is obtained for an observer with $\theta_{\rm v}>\theta_{\rm jet}+1/\bar{\Gamma}$.

In general, the observed spectral flux is expressed as ${f_\nu } \propto t_{{\rm{obs}}}^{ - \alpha }{\nu ^{ - \beta }}$ and the so-called ``closure relations'',
i.e., the relationship between the temporal index $\alpha$ and the spectral index $\beta$, are discussed
(e.g., \citealp{Zhang_B-2004-Meszaros_P-IJMPA.19.2385Z}; \citealp{Zhang_B-2006-Fan_YZ-ApJ.642.354Z}).
With Equations~(\ref{Eq:AP_Dynamics_On-axis}) and (\ref{Eq:AP_Dynamics_Off-axis}),
the values of $\alpha$ and $\beta$ for the situations with $\bar{R}>R_{\rm dec}$ and $\bar{\Gamma}\gtrsim 3$ are reported in the Table~\ref{MyTabB},
where the situations with $\bar{R}>R_{\rm dec}$
is related to the normal decay phase and the post-jet-break-phase of afterglows
if the GRB is observed on-axis.
The relations of $\alpha$ and $s$ for an off-axis observer can be found in Figure~\ref{Fig:alpha_s}.

\begin{table}
{\centering
\caption{The values of $\alpha$ and $\beta$ in different situations, where ${f_\nu } \propto t_{{\mathop{\rm obs}\nolimits} }^{ - \alpha }\nu^{-\beta}$ is adopted.}\label{MyTabB}
\begin{tabular}{c|c|c|c}
\hline\hline
Cases	& \multicolumn{3}{|c}{$\alpha$}\\
\hline																
Fast cooling & On-axis before the jet break & Free-wind medium &  Shocked-wind medium\\
\hline
${\nu  < {\nu _c}~(\beta=-\frac{1}{3})}$  &	$- \frac{{3 - 2s}}{\zeta } - 10\frac{{3 - s}}{{3\left( {\varepsilon  + 2s - 8} \right)}} - \frac{2}{3}$ & ${ \frac{2}{3}}$&${-\frac{1}{6}}$	\\
${{\nu _c} < \nu  < {\nu _m}~(\beta=\frac{1}{2})}$	& ${1-\frac{3}{\zeta }+\frac{3s}{4\zeta } }$&${ \frac{1}{4}}$&${ \frac{1}{4}}$\\
${\nu  > {\nu _m}~(\beta=\frac{p}{2})}$	&	${ -\frac{ {12 - 2s - sp} }{ {4\zeta }}- \frac{\left( {3 - s} \right)\left( {2p - 2} \right)}{{\varepsilon  + 2s - 8}} + 1}$  &  ${-\frac{2-3p}{4}}\sim 1.15$ &
 $-\frac{2-3p}{4}\sim 1.15$	\\
\hline																					
Slow cooling & On-axis before the jet break\\
\hline
$\nu  < {\nu _m}~(\beta=-\frac{1}{3})$	&	${-\frac{{2\left( {3 - s} \right)} }{{3\left( {\varepsilon  + 2s - 8} \right)}} - \frac{3}{\zeta }+\frac{4s}{3\zeta }}$	&${0}$&${-\frac{1}{2}}$\\
${\nu _m} < \nu  < {\nu _c}~(\beta=\frac{p-1}{2})$	&	${-\frac{2p\left( {3 - s} \right)}{ {\varepsilon  + 2s - 8} }+\frac{ s\left( {p - 1} \right)}{4\zeta}-
\frac{3}{\zeta}+\frac{3s}{2\zeta }}$&${-\frac{1-3p}{4}\sim 1.4}$&${-\frac{3(1-p)}{4} \sim 0.9}$	\\
$\nu  > {\nu _c}~(\beta=\frac{p}{2})$	&	$-\frac{\left( {3 - s} \right)\left( {2p - 2} \right)}{ {\varepsilon  + 2s - 8} }- \frac{3}{\zeta }+ \frac{s}{2\zeta }+ \frac{sp}{4\zeta }+1$	&
${-\frac{2-3p}{4}}\sim 1.15$&$-\frac{2-3p}{4}\sim 1.15$\\
\hline\hline																		
Fast cooling & Off-axis before the jet break\\
\hline
${\nu  < {\nu _c}}~(\beta=-\frac{1}{3})$	&	${-\frac{2\zeta }{3}- 3 +2s}$ &${-\frac{1}{3}}$&${-\frac{17}{3}}$	\\
${{\nu _c} < \nu  < {\nu _m}}~(\beta=\frac{1}{2})$	&	${  \frac{5\left( {3 - s} \right)}{ {\varepsilon  - 2}  }- 3 + \frac{3s}{4} + \zeta }$ &${-2}$&${-\frac{13}{2}}$	\\
${\nu  > {\nu _m}}~(\beta=\frac{p}{2})$	&	${-\frac{{12 - 2s - sp} }{4} - \frac{\left( {3 - s} \right)\left( {p - 6} \right)}{{\varepsilon  - 2} } + \zeta }$	&${-(3-p)}\sim -0.8$&
${-\frac{16-3p}{2}}\sim -4.7$\\
\hline																					
Slow cooling & Off-axis before the jet break\\
\hline
$\nu  < {\nu _m}~(\beta=-\frac{1}{3})$	&	${-3 + \frac{4s}{3} + \frac{8\left( {3 - s} \right)}{ {3\left( {\varepsilon  - 2} \right)} }}$	&${-\frac{5}{3}}$&${-7}$\\
${\nu _m} < \nu  < {\nu _c}~(\beta=\frac{p-1}{2})$	&	${  \frac{ s\left( {p + 1} \right)}{4} - 3 +s -\frac{\left( {3 - s} \right)\left( {p - 3} \right)}{\left( {\varepsilon  - 2} \right)}}$	&$-(2-p)\sim 0.2$&${-\frac{15-3p}{2}}\sim -4.2$\\
$\nu  > {\nu _c}~(\beta=\frac{p}{2})$	&	${-\frac{\left( {12 - 2s - sp} \right)}{4} - \frac{\left( {3 - s} \right)\left( {p - 6} \right)}{{\varepsilon  - 2} } +\zeta }$	&${-(3-p)}\sim -0.8$&${-\frac{16-3p}{2}}\sim -4.7$\\
\hline\hline																		
Fast cooling & Post-jet-break-phase \\
\hline
${\nu  < {\nu _c}}~(\beta=-\frac{1}{3})$  &	$ - \frac{{3 - 2s}}{\zeta } - 10\frac{{3 - s}}{{3\left( {\varepsilon  + 2s - 8} \right)}} - \frac{2}{3}-\frac{2(3-s)}{\varepsilon  + 2s - 8}$ & ${ \frac{7}{6}}$&${\frac{7}{12}}$	\\
${{\nu _c} < \nu  < {\nu _m}}~(\beta=\frac{1}{2})$	& ${1-\frac{3}{\zeta }+\frac{3s}{4\zeta } -\frac{2(3-s)}{\varepsilon  + 2s - 8}}$&${ \frac{3}{4}}$&${ 1}$\\
${\nu  > {\nu _m}}~(\beta=\frac{p}{2})$	&	${ -\frac{ {12 - 2s - sp} }{{4\zeta }  }- \frac{\left( {3 - s} \right)\left( {2p - 2} \right)}{{\varepsilon  + 2s - 8} } + 1-\frac{2(3-s)}{\varepsilon  + 2s - 8}}$  &  ${-\frac{2-3p}{4}}+\frac{1}{2}\sim 1.65$ & $-\frac{2-3p}{4}+\frac{3}{4}\sim 1.9$	\\
\hline																					
Slow cooling & Post-jet-break-phase \\
\hline
$\nu  < {\nu _m}~(\beta=-\frac{1}{3})$	&	${-\frac{{2\left( {3 - s} \right)} }{{3\left( {\varepsilon  + 2s - 8} \right)}} - \frac{3}{\zeta }+\frac{4s}{3\zeta }-\frac{2(3-s)}{\varepsilon  + 2s - 8}
}$	&${\frac{1}{2}}$&${\frac{1}{4}}$\\
${\nu _m} < \nu  < {\nu _c}~(\beta=\frac{p-1}{2})$	&	${-\frac{2p\left( {3 - s} \right)}{ {\varepsilon  + 2s - 8} }+\frac{ s\left( {p - 1} \right)}{4\zeta}-
\frac{3}{\zeta}+\frac{3s}{2\zeta }}-\frac{2(3-s)}{\varepsilon  + 2s - 8}$&${-\frac{1-3p}{4} +\frac{1}{2}\sim 1.9}$&${-\frac{3(1-p)}{4 }+\frac{3}{4}\sim 1.65}$	\\
$\nu  > {\nu _c}~(\beta=\frac{p}{2})$	&	$-\frac{\left( {3 - s} \right)\left( {2p - 2} \right)}{ {\varepsilon  + 2s - 8} }- \frac{3}{\zeta }+ \frac{s}{2\zeta }+ \frac{sp}{4\zeta }+1-\frac{2(3-s)}{\varepsilon  + 2s - 8}$	&${-\frac{2-3p}{4}} +\frac{1}{2}\sim 1.65$&$-\frac{2-3p}{4}+\frac{3}{4}\sim 1.9$\\
\hline																					
\end{tabular}\\}
\tablenotemark{}{\;\;\;\;\;\;\;\;\;\;\;\;\;\;\;\;\;\;$p=2.2$ is adopted.}
\end{table}

\begin{figure}
\centering
\includegraphics[angle=0,width=0.8\textwidth]{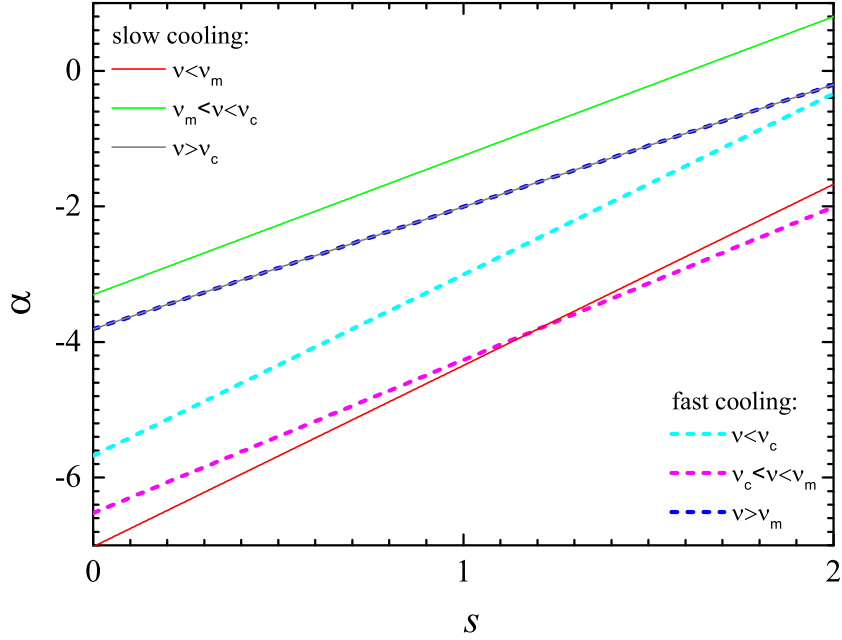}
\caption{Relation of $\alpha$ and $s$ for an off-axis observer and $\bar{R}>R_{\rm dec}$, where ${f_\nu } \propto t_{{\mathop{\rm obs}\nolimits} }^{ - \alpha }$ is adopted.
One can find that the rise of the free-wind-afterglow is generally gentle compared with that of the shocked-wind-afterglow.}\label{Fig:alpha_s}
\end{figure}
}
\clearpage

\begin{thebibliography}{}
\expandafter\ifx\csname natexlab\endcsname\relax\def\natexlab#1{#1}\fi

\bibitem[{{Barthelmy} {et~al.}(2005){Barthelmy}, {Cannizzo}, {Gehrels},
  {Cusumano}, {Mangano}, {O'Brien}, {Vaughan}, {Zhang}, {Burrows}, {Campana},
  {Chincarini}, {Goad}, {Kouveliotou}, {Kumar}, {M{\'e}sz{\'a}ros}, {Nousek},
  {Osborne}, {Panaitescu}, {Reeves}, {Sakamoto}, {Tagliaferri}, \&
  {Wijers}}]{Barthelmy_SD-2005-Cannizzo_JK-ApJ.635L.133B}
{Barthelmy}, S.~D., {Cannizzo}, J.~K., {Gehrels}, N., {et~al.} 2005, \apjl,
  635, L133

\bibitem[{{Burrows} {et~al.}(2005{\natexlab{a}}){Burrows}, {Romano}, {Falcone},
  {Kobayashi}, {Zhang}, {Moretti}, {O'Brien}, {Goad}, {Campana}, {Page},
  {Angelini}, {Barthelmy}, {Beardmore}, {Capalbi}, {Chincarini}, {Cummings},
  {Cusumano}, {Fox}, {Giommi}, {Hill}, {Kennea}, {Krimm}, {Mangano},
  {Marshall}, {M{\'e}sz{\'a}ros}, {Morris}, {Nousek}, {Osborne}, {Pagani},
  {Perri}, {Tagliaferri}, {Wells}, {Woosley}, \&
  {Gehrels}}]{Burrows_DN-2005-Romano_P-Sci.309.1833B}
{Burrows}, D.~N., {Romano}, P., {Falcone}, A., {et~al.} 2005{\natexlab{a}},
  Science, 309, 1833

\bibitem[{{Burrows} {et~al.}(2005{\natexlab{b}}){Burrows}, {Hill}, {Nousek},
  {Kennea}, {Wells}, {Osborne}, {Abbey}, {Beardmore}, {Mukerjee}, {Short},
  {Chincarini}, {Campana}, {Citterio}, {Moretti}, {Pagani}, {Tagliaferri},
  {Giommi}, {Capalbi}, {Tamburelli}, {Angelini}, {Cusumano}, {Br{\"a}uninger},
  {Burkert}, \& {Hartner}}]{Burrows_DN-2005-Hill_JE-SSRv.120.165B}
{Burrows}, D.~N., {Hill}, J.~E., {Nousek}, J.~A., {et~al.} 2005{\natexlab{b}},
  \ssr, 120, 165

\bibitem[{{Cannizzo} {et~al.}(2004){Cannizzo}, {Gehrels}, \&
  {Vishniac}}]{Cannizzo_JK-2004-Gehrels_N-ApJ.601.380C}
{Cannizzo}, J.~K., {Gehrels}, N., \& {Vishniac}, E.~T. 2004, \apj, 601, 380

\bibitem[{{Castor} {et~al.}(1975){Castor}, {McCray}, \&
  {Weaver}}]{Castor_J-1975-McCray_R-ApJ.200L.107C}
{Castor}, J., {McCray}, R., \& {Weaver}, R. 1975, \apjl, 200, L107

\bibitem[{{Chevalier} \& {Li}(2000)}]{Chevalier_RA-2000-Li_ZY-ApJ.536.195C}
{Chevalier}, R.~A., \& {Li}, Z.-Y. 2000, \apj, 536, 195

\bibitem[{{Chevalier} {et~al.}(2004){Chevalier}, {Li}, \&
  {Fransson}}]{Chevalier_RA-2004-Li_ZY-ApJ.606.369C}
{Chevalier}, R.~A., {Li}, Z.-Y., \& {Fransson}, C. 2004, \apj, 606, 369

\bibitem[{{Chiang} \& {Dermer}(1999)}]{Chiang_J-1999-Dermer_CD-ApJ.512.699C}
{Chiang}, J., \& {Dermer}, C.~D. 1999, \apj, 512, 699

\bibitem[{{Chincarini} {et~al.}(2007){Chincarini}, {Moretti}, {Romano},
  {Falcone}, {Morris}, {Racusin}, {Campana}, {Covino}, {Guidorzi},
  {Tagliaferri}, {Burrows}, {Pagani}, {Stroh}, {Grupe}, {Capalbi}, {Cusumano},
  {Gehrels}, {Giommi}, {La Parola}, {Mangano}, {Mineo}, {Nousek}, {O'Brien},
  {Page}, {Perri}, {Troja}, {Willingale}, \&
  {Zhang}}]{Chincarini_G-2007-Moretti_A-ApJ.671.1903C}
{Chincarini}, G., {Moretti}, A., {Romano}, P., {et~al.} 2007, \apj, 671, 1903

\bibitem[{{Chincarini} {et~al.}(2010){Chincarini}, {Mao}, {Margutti},
  {Bernardini}, {Guidorzi}, {Pasotti}, {Giannios}, {Della Valle}, {Moretti},
  {Romano}, {D'Avanzo}, {Cusumano}, \&
  {Giommi}}]{Chincarini_G-2010-Mao_J-MNRAS.406.2113C}
{Chincarini}, G., {Mao}, J., {Margutti}, R., {et~al.} 2010, \mnras, 406, 2113

\bibitem[{{De Pasquale} {et~al.}(2015){De Pasquale}, {Kuin}, {Oates},
  {Schulze}, {Cano}, {Guidorzi}, {Beardmore}, {Evans}, {Uhm}, {Zhang}, {Page},
  {Kobayashi}, {Castro-Tirado}, {Gorosabel}, {Sakamoto}, {Fatkhullin},
  {Pandey}, {Im}, {Chandra}, {Frail}, {Gao}, {Kopa{\v{c}}}, {Jeon}, {Akerlof},
  {Huang}, {Pak}, {Park}, {Gomboc}, {Melandri}, {Zane}, {Mundell}, {Saxton},
  {Holland}, {Virgili}, {Urata}, {Steele}, {Bersier}, {Tanvir}, {Sokolov}, \&
  {Moskvitin}}]{DePasquale_Massimiliano-2015-Kuin_NPM-MNRAS.449.1024}
{De Pasquale}, M., {Kuin}, N.~P.~M., {Oates}, S., {et~al.} 2015, \mnras, 449,
  1024

\bibitem[{{Dyson} \& {Williams}(1997)}]{Dyson_JE-1997-Williams_DA-pism.book.D}
{Dyson}, J.~E., \& {Williams}, D.~A. 1997, {The physics of the interstellar
  medium}, doi:10.1201/9780585368115

\bibitem[{{Falcone} {et~al.}(2006){Falcone}, {Burrows}, {Lazzati}, {Campana},
  {Kobayashi}, {Zhang}, {M{\'e}sz{\'a}ros}, {Page}, {Kennea}, {Romano},
  {Pagani}, {Angelini}, {Beardmore}, {Capalbi}, {Chincarini}, {Cusumano},
  {Giommi}, {Goad}, {Godet}, {Grupe}, {Hill}, {La Parola}, {Mangano},
  {Moretti}, {Nousek}, {O'Brien}, {Osborne}, {Perri}, {Tagliaferri}, {Wells},
  \& {Gehrels}}]{Falcone_AD-2006-Burrows_DN-ApJ.641.1010F}
{Falcone}, A.~D., {Burrows}, D.~N., {Lazzati}, D., {et~al.} 2006, \apj, 641,
  1010

\bibitem[{{Falcone} {et~al.}(2007){Falcone}, {Morris}, {Racusin}, {Chincarini},
  {Moretti}, {Romano}, {Burrows}, {Pagani}, {Stroh}, {Grupe}, {Campana},
  {Covino}, {Tagliaferri}, {Willingale}, \&
  {Gehrels}}]{Falcone_AD-2007-Morris_D-ApJ.671.1921F}
{Falcone}, A.~D., {Morris}, D., {Racusin}, J., {et~al.} 2007, \apj, 671, 1921

\bibitem[{{Feng} \&
  {Dai}(2011{\natexlab{a}})}]{Feng_Si-Yi-2011-Dai_Zi-Gao-RAA.11.1046F}
{Feng}, S.-Y., \& {Dai}, Z.-G. 2011{\natexlab{a}}, Research in Astronomy and
  Astrophysics, 11, 1046

\bibitem[{{Feng} \&
  {Dai}(2011{\natexlab{b}})}]{Feng_SY-2011-Dai_ZG-RAA.11.1046F}
---. 2011{\natexlab{b}}, Research in Astronomy and Astrophysics, 11, 1046

\bibitem[{{Fraija} {et~al.}(2020){Fraija}, {De Colle}, {Veres}, {Dichiara},
  {Barniol Duran}, {Caligula do E.~S. Pedreira}, {Galvan-Gamez}, \& {Betancourt
  Kamenetskaia}}]{Fraija_N-2020-DeColle_F-ApJ.896.25F}
{Fraija}, N., {De Colle}, F., {Veres}, P., {et~al.} 2020, \apj, 896, 25

\bibitem[{{Fraija} {et~al.}(2019){Fraija}, {Dichiara}, {Pedreira},
  {Galvan-Gamez}, {Becerra}, {Barniol Duran}, \&
  {Zhang}}]{Fraija_N-2019-Dichiara_S-ApJ.879L.26F}
{Fraija}, N., {Dichiara}, S., {Pedreira}, A.~C. C. d. E.~S., {et~al.} 2019,
  \apjl, 879, L26

\bibitem[{{Fraija} {et~al.}(2017){Fraija}, {Veres}, {Zhang}, {Barniol Duran},
  {Becerra}, {Zhang}, {Lee}, {Watson}, {Ordaz-Salazar}, \&
  {Galvan-Gamez}}]{Fraija_N-2017-Veres_P-ApJ.848.15F}
{Fraija}, N., {Veres}, P., {Zhang}, B.~B., {et~al.} 2017, \apj, 848, 15

\bibitem[{{Garcia-Segura} {et~al.}(1996){Garcia-Segura}, {Langer}, \& {Mac
  Low}}]{GarciaSegura_G-1996-Langer_N-A&A.316.133G}
{Garcia-Segura}, G., {Langer}, N., \& {Mac Low}, M.~M. 1996, \aap, 316, 133

\bibitem[{{Gehrels} {et~al.}(2004){Gehrels}, {Chincarini}, {Giommi}, {Mason},
  {Nousek}, {Wells}, {White}, {Barthelmy}, {Burrows}, {Cominsky}, {Hurley},
  {Marshall}, {M{\'e}sz{\'a}ros}, {Roming}, {Angelini}, {Barbier}, {Belloni},
  {Campana}, {Caraveo}, {Chester}, {Citterio}, {Cline}, {Cropper}, {Cummings},
  {Dean}, {Feigelson}, {Fenimore}, {Frail}, {Fruchter}, {Garmire}, {Gendreau},
  {Ghisellini}, {Greiner}, {Hill}, {Hunsberger}, {Krimm}, {Kulkarni}, {Kumar},
  {Lebrun}, {Lloyd-Ronning}, {Markwardt}, {Mattson}, {Mushotzky}, {Norris},
  {Osborne}, {Paczynski}, {Palmer}, {Park}, {Parsons}, {Paul}, {Rees},
  {Reynolds}, {Rhoads}, {Sasseen}, {Schaefer}, {Short}, {Smale}, {Smith},
  {Stella}, {Tagliaferri}, {Takahashi}, {Tashiro}, {Townsley}, {Tueller},
  {Turner}, {Vietri}, {Voges}, {Ward}, {Willingale}, {Zerbi}, \&
  {Zhang}}]{Gehrels_N-2004-Chincarini_G-ApJ.611.1005G}
{Gehrels}, N., {Chincarini}, G., {Giommi}, P., {et~al.} 2004, \apj, 611, 1005

\bibitem[{{Geng} {et~al.}(2018){Geng}, {Huang}, {Wu}, {Zhang}, \&
  {Zong}}]{Geng_JJ-2018-Huang_YF-ApJS.234.3G}
{Geng}, J.-J., {Huang}, Y.-F., {Wu}, X.-F., {Zhang}, B., \& {Zong}, H.-S. 2018,
  \apjs, 234, 3

\bibitem[{{Geng} {et~al.}(2016){Geng}, {Wu}, {Huang}, {Li}, \&
  {Dai}}]{Geng_JJ-2016-Wu_XF-ApJ.825.107G}
{Geng}, J.~J., {Wu}, X.~F., {Huang}, Y.~F., {Li}, L., \& {Dai}, Z.~G. 2016,
  \apj, 825, 107

\bibitem[{{Gorbovskoy} {et~al.}(2012){Gorbovskoy}, {Lipunova}, {Lipunov},
  {Kornilov}, {Belinski}, {Shatskiy}, {Tyurina}, {Kuvshinov}, {Balanutsa},
  {Chazov}, {Kuznetsov}, {Zimnukhov}, {Kornilov}, {Sankovich}, {Krylov},
  {Ivanov}, {Chvalaev}, {Poleschuk}, {Konstantinov}, {Gress}, {Yazev},
  {Budnev}, {Krushinski}, {Zalozhnich}, {Popov}, {Tlatov}, {Parhomenko},
  {Dormidontov}, {Senik}, {Yurkov}, {Sergienko}, {Varda}, {Kudelina},
  {Castro-Tirado}, {Gorosabel}, {S{\'a}nchez-Ram{\'{\i}}rez}, {Jelinek}, \&
  {Tello}}]{Gorbovskoy_ES-2012-Lipunova_GV-MNRAS.421.1874G}
{Gorbovskoy}, E.~S., {Lipunova}, G.~V., {Lipunov}, V.~M., {et~al.} 2012,
  \mnras, 421, 1874

\bibitem[{{Granot} {et~al.}(2018){Granot}, {Gill}, {Guetta}, \& {De
  Colle}}]{Granot_J-2018-Gill_R-MNRAS.481.1597G}
{Granot}, J., {Gill}, R., {Guetta}, D., \& {De Colle}, F. 2018, \mnras, 481,
  1597

\bibitem[{{Granot} {et~al.}(2001){Granot}, {Miller}, {Piran}, {Suen}, \&
  {Hughes}}]{Granot_J-2001-Miller_M-grba.conf.312G}
{Granot}, J., {Miller}, M., {Piran}, T., {Suen}, W.~M., \& {Hughes}, P.~A.
  2001, in Gamma-ray Bursts in the Afterglow Era, ed. E.~{Costa},
  F.~{Frontera}, \& J.~{Hjorth}, 312

\bibitem[{{Guidorzi} {et~al.}(2014){Guidorzi}, {Mundell}, {Harrison},
  {Margutti}, {Sudilovsky}, {Zauderer}, {Kobayashi}, {Cucchiara}, {Melandri},
  {Pandey}, {Berger}, {Bersier}, {D'Elia}, {Gomboc}, {Greiner}, {Japelj},
  {Kopa{\v c}}, {Kumar}, {Malesani}, {Mottram}, {O'Brien}, {Rau}, {Smith},
  {Steele}, {Tanvir}, \& {Virgili}}]{Guidorzi_C-2014-Mundell_CG-MNRAS.438.752G}
{Guidorzi}, C., {Mundell}, C.~G., {Harrison}, R., {et~al.} 2014, \mnras, 438,
  752

\bibitem[{{Hou} {et~al.}(2014){Hou}, {Geng}, {Wang}, {Wu}, {Huang}, {Dai}, \&
  {Lu}}]{Hou_SJ-2014-Geng_JJ-ApJ.785.113H}
{Hou}, S.~J., {Geng}, J.~J., {Wang}, K., {et~al.} 2014, \apj, 785, 113

\bibitem[{{Huang} {et~al.}(2019){Huang}, {Lin}, {Liu}, {Ren}, {Wang}, {Liu}, \&
  {Liang}}]{Huang_BQ-2019-Lin_DB-MNRAS.487.3214H}
{Huang}, B.-Q., {Lin}, D.-B., {Liu}, T., {et~al.} 2019, \mnras, 487, 3214

\bibitem[{{Huang} {et~al.}(2018){Huang}, {Wang}, {Zheng}, {Liang}, {Lin},
  {Zhong}, {Zhang}, {Huang}, {Filippenko}, \&
  {Zhang}}]{Huang_LY-2018-Wang_XG-ApJ.859.163H}
{Huang}, L.-Y., {Wang}, X.-G., {Zheng}, W., {et~al.} 2018, \apj, 859, 163

\bibitem[{{Huang} {et~al.}(1999){Huang}, {Dai}, \&
  {Lu}}]{Huang_YF-1999-Dai_ZG-MNRAS.309.513H}
{Huang}, Y.~F., {Dai}, Z.~G., \& {Lu}, T. 1999, \mnras, 309, 513

\bibitem[{{Huang} {et~al.}(2000){Huang}, {Gou}, {Dai}, \&
  {Lu}}]{Huang_YF-2000-Gou_LJ-ApJ.543.90H}
{Huang}, Y.~F., {Gou}, L.~J., {Dai}, Z.~G., \& {Lu}, T. 2000, \apj, 543, 90

\bibitem[{{Jin} {et~al.}(2009{\natexlab{a}}){Jin}, {Xu}, {Covino}, {D'Avanzo},
  {Antonelli}, {Fan}, \& {Wei}}]{Jin_ZP-2009-Xu_D-MNRAS.400.1829J}
{Jin}, Z.~P., {Xu}, D., {Covino}, S., {et~al.} 2009{\natexlab{a}}, \mnras, 400,
  1829

\bibitem[{{Jin} {et~al.}(2009{\natexlab{b}}){Jin}, {Xu}, {Covino}, {D'Avanzo},
  {Antonelli}, {Fan}, \& {Wei}}]{Jin_ZP-2009-Xu_D_MNRAS.400.1829J}
---. 2009{\natexlab{b}}, \mnras, 400, 1829

\bibitem[{{Kamble} {et~al.}(2007){Kamble}, {Resmi}, \&
  {Misra}}]{Kamble_Atish-2007-Resmi_L-ApJ.664L.5K}
{Kamble}, A., {Resmi}, L., \& {Misra}, K. 2007, \apjl, 664, L5

\bibitem[{{Kong} {et~al.}(2010){Kong}, {Wong}, {Huang}, \&
  {Cheng}}]{Kong_SW-2010-Wong_AYL-MNRAS.402.409K}
{Kong}, S.~W., {Wong}, A.~Y.~L., {Huang}, Y.~F., \& {Cheng}, K.~S. 2010,
  \mnras, 402, 409

\bibitem[{{Kumar} {et~al.}(2012){Kumar}, {Hern{\'a}ndez}, {Bo{\v{s}}njak}, \&
  {Barniol Duran}}]{Kumar_P-2012-Hernandez_RA-MNRAS.427L.40K}
{Kumar}, P., {Hern{\'a}ndez}, R.~A., {Bo{\v{s}}njak}, {\v{Z}}., \& {Barniol
  Duran}, R. 2012, \mnras, 427, L40

\bibitem[{{Lamb} {et~al.}(2019){Lamb}, {Lyman}, {Levan}, {Tanvir}, {Kangas},
  {Fruchter}, {Gompertz}, {Hjorth}, {Mandel}, {Oates}, {Steeghs}, \&
  {Wiersema}}]{Lamb_GP-2019-Lyman_JD-ApJ.870L.15L}
{Lamb}, G.~P., {Lyman}, J.~D., {Levan}, A.~J., {et~al.} 2019, \apj, 870, L15

\bibitem[{{Laskar} {et~al.}(2015){Laskar}, {Berger}, {Margutti}, {Perley},
  {Zauderer}, {Sari}, \& {Fong}}]{Laskar_T-2015-Berger_E-ApJ.814.1L}
{Laskar}, T., {Berger}, E., {Margutti}, R., {et~al.} 2015, \apj, 814, 1

\bibitem[{{Li} {et~al.}(2012){Li}, {Liang}, {Tang}, {Chen}, {Xi}, {L{\"u}},
  {Gao}, {Zhang}, {Zhang}, {Yi}, {Lu}, {L{\"u}}, \&
  {Wei}}]{Li_L-2012-Liang_EW-ApJ.758.27L}
{Li}, L., {Liang}, E.-W., {Tang}, Q.-W., {et~al.} 2012, \apj, 758, 27

\bibitem[{{Li} {et~al.}(2020){Li}, {Wang}, {Zheng}, {Pozanenko}, {Filippenko},
  {Qin}, {Wang}, {Jiang}, {Li}, {Lin}, {Liang}, {Volnova}, {Elenin}, {Klunko},
  {Inasaridze}, {Kusakin}, \& {Lu}}]{Li_L-2020-Wang_XG-ApJ.900.176L}
{Li}, L., {Wang}, X.-G., {Zheng}, W., {et~al.} 2020, \apj, 900, 176

\bibitem[{{Liang} {et~al.}(2006){Liang}, {Zhang}, {O'Brien}, {Willingale},
  {Angelini}, {Burrows}, {Campana}, {Chincarini}, {Falcone}, {Gehrels}, {Goad},
  {Grupe}, {Kobayashi}, {M{\'e}sz{\'a}ros}, {Nousek}, {Osborne}, {Page}, \&
  {Tagliaferri}}]{Liang_EW-2006-Zhang_B-ApJ.646.351L}
{Liang}, E.~W., {Zhang}, B., {O'Brien}, P.~T., {et~al.} 2006, \apj, 646, 351

\bibitem[{{Liang} {et~al.}(2013){Liang}, {Li}, {Gao}, {Zhang}, {Liang}, {Wu},
  {Yi}, {Dai}, {Tang}, {Chen}, {L{\"u}}, {Zhang}, {Lu}, {L{\"u}}, \&
  {Wei}}]{Liang_EW-2013-Li_L-ApJ.774.13L}
{Liang}, E.-W., {Li}, L., {Gao}, H., {et~al.} 2013, \apj, 774, 13

\bibitem[{{Lin} {et~al.}(2017{\natexlab{a}}){Lin}, {Mu}, {Liang}, {Liu}, {Gu},
  {Lu}, {Wang}, \& {Liang}}]{Lin_DB-2017-Mu_HJ-ApJ.840.118L}
{Lin}, D.-B., {Mu}, H.-J., {Liang}, Y.-F., {et~al.} 2017{\natexlab{a}}, \apj,
  840, 118

\bibitem[{{Lin} {et~al.}(2017{\natexlab{b}}){Lin}, {Mu}, {Lu}, {Liu}, {Gu},
  {Liang}, {Wang}, \& {Liang}}]{Lin_DB-2017-Mu_HJ-ApJ.840.95L}
{Lin}, D.-B., {Mu}, H.-J., {Lu}, R.-J., {et~al.} 2017{\natexlab{b}}, \apj, 840,
  95

\bibitem[{{Marshall} {et~al.}(2011){Marshall}, {Antonelli}, {Burrows},
  {Covino}, {de Pasquale}, {Evans}, {Fugazza}, {Holland}, {Liang}, {O'Brien},
  {Oates}, {Osborne}, {Pagani}, {Sakamoto}, {Siegel}, {Wu}, \&
  {Zhang}}]{Marshall_FE-2011-Antonelli_LA-ApJ.727.132M}
{Marshall}, F.~E., {Antonelli}, L.~A., {Burrows}, D.~N., {et~al.} 2011, \apj,
  727, 132

\bibitem[{{Melandri} {et~al.}(2014){Melandri}, {Virgili}, {Guidorzi},
  {Bernardini}, {Kobayashi}, {Mundell}, {Gomboc}, {Dintinjana}, {Hentunen},
  {Japelj}, {Kopa{\v{c}}}, {Kuroda}, {Morgan}, {Steele}, {Quadri}, {Arici},
  {Arnold}, {Girelli}, {Hanayama}, {Kawai}, {Miku{\v{z}}}, {Nissinen}, {Salmi},
  {Smith}, {Strabla}, {Tonincelli}, \&
  {Quadri}}]{Melandri_A-2014-Virgili_FJ-A&A.572A.55M}
{Melandri}, A., {Virgili}, F.~J., {Guidorzi}, C., {et~al.} 2014, \aap, 572, A55

\bibitem[{{M{\'e}sz{\'a}ros} \&
  {Rees}(1999)}]{Meszaros_P-1999-Rees_MJ-MNRAS.306L.39M}
{M{\'e}sz{\'a}ros}, P., \& {Rees}, M.~J. 1999, \mnras, 306, L39

\bibitem[{{Mu} {et~al.}(2016{\natexlab{a}}){Mu}, {Gu}, {Hou}, {Liu}, {Lin},
  {Yi}, {Liang}, \& {Lu}}]{Mu_HJ-2016-Gu_WM-ApJ.832.161M}
{Mu}, H.-J., {Gu}, W.-M., {Hou}, S.-J., {et~al.} 2016{\natexlab{a}}, \apj, 832,
  161

\bibitem[{{Mu} {et~al.}(2016{\natexlab{b}}){Mu}, {Lin}, {Xi}, {Lin}, {Wang},
  {Liang}, {L{\"u}}, {Zhang}, \& {Liang}}]{Mu_HJ-2016-Lin_DB-ApJ.831.111M}
{Mu}, H.-J., {Lin}, D.-B., {Xi}, S.-Q., {et~al.} 2016{\natexlab{b}}, \apj, 831,
  111

\bibitem[{{Nardini} {et~al.}(2014){Nardini}, {Elliott}, {Filgas}, {Schady},
  {Greiner}, {Kr{\"u}hler}, {Klose}, {Afonso}, {Kann}, {Nicuesa Guelbenzu},
  {Olivares E.}, {Rau}, {Rossi}, {Sudilovsky}, \&
  {Schmidl}}]{Nardini_M-2014-Elliott_J-A&A.562A.29N}
{Nardini}, M., {Elliott}, J., {Filgas}, R., {et~al.} 2014, \aap, 562, A29

\bibitem[{{Nousek} {et~al.}(2006){Nousek}, {Kouveliotou}, {Grupe}, {Page},
  {Granot}, {Ramirez-Ruiz}, {Patel}, {Burrows}, {Mangano}, {Barthelmy},
  {Beardmore}, {Campana}, {Capalbi}, {Chincarini}, {Cusumano}, {Falcone},
  {Gehrels}, {Giommi}, {Goad}, {Godet}, {Hurkett}, {Kennea}, {Moretti},
  {O'Brien}, {Osborne}, {Romano}, {Tagliaferri}, \&
  {Wells}}]{Nousek_JA-2006-Kouveliotou_C-ApJ.642.389N}
{Nousek}, J.~A., {Kouveliotou}, C., {Grupe}, D., {et~al.} 2006, \apj, 642, 389

\bibitem[{{O'Brien} {et~al.}(2006){O'Brien}, {Willingale}, {Osborne}, {Goad},
  {Page}, {Vaughan}, {Rol}, {Beardmore}, {Godet}, {Hurkett}, {Wells}, {Zhang},
  {Kobayashi}, {Burrows}, {Nousek}, {Kennea}, {Falcone}, {Grupe}, {Gehrels},
  {Barthelmy}, {Cannizzo}, {Cummings}, {Hill}, {Krimm}, {Chincarini},
  {Tagliaferri}, {Campana}, {Moretti}, {Giommi}, {Perri}, {Mangano}, \&
  {LaParola}}]{OBrien_PT-2006-Willingale_R-ApJ.647.1213O}
{O'Brien}, P.~T., {Willingale}, R., {Osborne}, J., {et~al.} 2006, \apj, 647,
  1213

\bibitem[{{Panaitescu} {et~al.}(2006){Panaitescu}, {M{\'e}sz{\'a}ros},
  {Burrows}, {Nousek}, {Gehrels}, {O'Brien}, \&
  {Willingale}}]{Panaitescu_A-2006-Meszaros_P-MNRAS.369.2059P}
{Panaitescu}, A., {M{\'e}sz{\'a}ros}, P., {Burrows}, D., {et~al.} 2006, \mnras,
  369, 2059

\bibitem[{{Panaitescu} \&
  {Vestrand}(2011)}]{Panaitescu_A-2011-Vestrand_WT-MNRAS.414.3537P}
{Panaitescu}, A., \& {Vestrand}, W.~T. 2011, \mnras, 414, 3537

\bibitem[{{Pe'er} \& {Wijers}(2006)}]{Peer_A-2006-Wijers_RAMJ-ApJ.643.1036P}
{Pe'er}, A., \& {Wijers}, R. A.~M.~J. 2006, \apj, 643, 1036

\bibitem[{{Piran}(1999)}]{Piran_T-1999-PhR.314.575P}
{Piran}, T. 1999, \physrep, 314, 575

\bibitem[{{Racusin} {et~al.}(2011){Racusin}, {Oates}, {Schady}, {Burrows}, {de
  Pasquale}, {Donato}, {Gehrels}, {Koch}, {McEnery}, {Piran}, {Roming},
  {Sakamoto}, {Swenson}, {Troja}, {Vasileiou}, {Virgili}, {Wanderman}, \&
  {Zhang}}]{Racusin_JL-2011-Oates_SR-ApJ.738.138R}
{Racusin}, J.~L., {Oates}, S.~R., {Schady}, P., {et~al.} 2011, \apj, 738, 138

\bibitem[{{Ramirez-Ruiz} {et~al.}(2001){Ramirez-Ruiz}, {Dray}, {Madau}, \&
  {Tout}}]{Ramirez-Ruiz_Enrico-2001-Dray_LynnetteM-MNRAS.327.829R}
{Ramirez-Ruiz}, E., {Dray}, L.~M., {Madau}, P., \& {Tout}, C.~A. 2001, \mnras,
  327, 829

\bibitem[{{Ren} {et~al.}(2020){Ren}, {Lin}, {Zhang}, {Wang}, {Li}, {Wang}, \&
  {Liang}}]{Ren_J-2020-Lin_DB-ApJ.901L.26R}
{Ren}, J., {Lin}, D.-B., {Zhang}, L.-L., {et~al.} 2020, \apjl, 901, L26

\bibitem[{{Romano} {et~al.}(2006){Romano}, {Moretti}, {Banat}, {Burrows},
  {Campana}, {Chincarini}, {Covino}, {Malesani}, {Tagliaferri}, {Kobayashi},
  {Zhang}, {Falcone}, {Angelini}, {Barthelmy}, {Beardmore}, {Capalbi},
  {Cusumano}, {Giommi}, {Goad}, {Godet}, {Grupe}, {Hill}, {Kennea}, {La
  Parola}, {Mangano}, {M{\'e}sz{\'a}ros}, {Morris}, {Nousek}, {O'Brien},
  {Osborne}, {Parsons}, {Perri}, {Pagani}, {Page}, {Wells}, \&
  {Gehrels}}]{Romano_P-2006-Moretti_A-A&A.450.59R}
{Romano}, P., {Moretti}, A., {Banat}, P.~L., {et~al.} 2006, \aap, 450, 59

\bibitem[{{Sari} \&
  {Piran}(1999{\natexlab{a}})}]{Sari_R-1999-Piran_T-ApJ.517L.109S}
{Sari}, R., \& {Piran}, T. 1999{\natexlab{a}}, \apjl, 517, L109

\bibitem[{{Sari} \&
  {Piran}(1999{\natexlab{b}})}]{Sari_R-1999-Piran_T-ApJ.520.641S}
---. 1999{\natexlab{b}}, \apj, 520, 641

\bibitem[{{Sari} {et~al.}(1998){Sari}, {Piran}, \&
  {Narayan}}]{Sari_R-1998-Piran_T-ApJ.497L.17S}
{Sari}, R., {Piran}, T., \& {Narayan}, R. 1998, \apjl, 497, L17

\bibitem[{{Scalo} \&
  {Wheeler}(2001)}]{Scalo_John-2001-Wheeler_JCraig-ApJ.562.664S}
{Scalo}, J., \& {Wheeler}, J.~C. 2001, \apj, 562, 664

\bibitem[{{Troja} {et~al.}(2018){Troja}, {Piro}, {Ryan}, {van Eerten}, {Ricci},
  {Wieringa}, {Lotti}, {Sakamoto}, \&
  {Cenko}}]{Troja_E-2018-Piro_L-MNRAS.478L.18T}
{Troja}, E., {Piro}, L., {Ryan}, G., {et~al.} 2018, \mnras, 478, L18

\bibitem[{{Troja} {et~al.}(2019){Troja}, {van Eerten}, {Ryan}, {Ricci},
  {Burgess}, {Wieringa}, {Piro}, {Cenko}, \&
  {Sakamoto}}]{Troja_E-2019-vanEerten_H-MNRAS.tmp.2169T}
{Troja}, E., {van Eerten}, H., {Ryan}, G., {et~al.} 2019, \mnras, 2169

\bibitem[{{Urata} {et~al.}(2014){Urata}, {Huang}, {Takahashi}, {Im}, {Yamaoka},
  {Tashiro}, {Kim}, {Jang}, \& {Pak}}]{Urata_Y-2014-Huang_K-ApJ.789.146U}
{Urata}, Y., {Huang}, K., {Takahashi}, S., {et~al.} 2014, \apj, 789, 146

\bibitem[{{van Eerten} {et~al.}(2010){van Eerten}, {Zhang}, \&
  {MacFadyen}}]{vanEerten_H-2010-Zhang_W-ApJ.722.235V}
{van Eerten}, H., {Zhang}, W., \& {MacFadyen}, A. 2010, \apj, 722, 235

\bibitem[{{van Eerten} \&
  {MacFadyen}(2012)}]{vanEerten_HJ-2012-MacFadyen_AI-ApJ.751.155V}
{van Eerten}, H.~J., \& {MacFadyen}, A.~I. 2012, \apj, 751, 155

\bibitem[{{Wang} {et~al.}(2015){Wang}, {Zhang}, {Liang}, {Gao}, {Li}, {Deng},
  {Qin}, {Tang}, {Kann}, {Ryde}, \& {Kumar}}]{Wang_XG-2015-Zhang_B-ApJS.219.9W}
{Wang}, X.-G., {Zhang}, B., {Liang}, E.-W., {et~al.} 2015, \apjs, 219, 9

\bibitem[{{Weaver} {et~al.}(1977){Weaver}, {McCray}, {Castor}, {Shapiro}, \&
  {Moore}}]{Weaver_R-1977-McCray_R-ApJ.218.377W}
{Weaver}, R., {McCray}, R., {Castor}, J., {Shapiro}, P., \& {Moore}, R. 1977,
  \apj, 218, 377

\bibitem[{{Wu} {et~al.}(2013){Wu}, {Hou}, \&
  {Lei}}]{Wu_XF-2013-Hou_SJ-ApJ.767L.36W}
{Wu}, X.-F., {Hou}, S.-J., \& {Lei}, W.-H. 2013, \apjl, 767, L36

\bibitem[{{Yi} {et~al.}(2017){Yi}, {Lei}, {Zhang}, {Dai}, {Wu}, \&
  {Liang}}]{Yi_SX-2017-Lei_WH-JHEAp.13.1Y}
{Yi}, S.-X., {Lei}, W.-H., {Zhang}, B., {et~al.} 2017, Journal of High Energy
  Astrophysics, 13, 1

\bibitem[{{Yi} {et~al.}(2015){Yi}, {Wu}, {Wang}, \&
  {Dai}}]{Yi_SX-2015-Wu_XF-ApJ.807.92Y}
{Yi}, S.-X., {Wu}, X.-F., {Wang}, F.-Y., \& {Dai}, Z.-G. 2015, \apj, 807, 92

\bibitem[{{Yi} {et~al.}(2016){Yi}, {Xi}, {Yu}, {Wang}, {Mu}, {L{\"u}}, \&
  {Liang}}]{Yi_SX-2016-Xi_SQ-ApJS.224.20Y}
{Yi}, S.-X., {Xi}, S.-Q., {Yu}, H., {et~al.} 2016, \apjs, 224, 20

\bibitem[{{Yu} {et~al.}(2015){Yu}, {Huang}, {Wu}, {Xu}, \&
  {Geng}}]{Yu_YB-2015-Huang_YF-ApJ.805.88Y}
{Yu}, Y.~B., {Huang}, Y.~F., {Wu}, X.~F., {Xu}, M., \& {Geng}, J.~J. 2015,
  \apj, 805, 88

\bibitem[{{Zhang} {et~al.}(2006){Zhang}, {Fan}, {Dyks}, {Kobayashi},
  {M{\'e}sz{\'a}ros}, {Burrows}, {Nousek}, \&
  {Gehrels}}]{Zhang_B-2006-Fan_YZ-ApJ.642.354Z}
{Zhang}, B., {Fan}, Y.~Z., {Dyks}, J., {et~al.} 2006, \apj, 642, 354

\bibitem[{{Zhang} \&
  {M{\'e}sz{\'a}ros}(2004)}]{Zhang_B-2004-Meszaros_P-IJMPA.19.2385Z}
{Zhang}, B., \& {M{\'e}sz{\'a}ros}, P. 2004, International Journal of Modern
  Physics A, 19, 2385

\bibitem[{{Zhang} {et~al.}(2007){Zhang}, {Liang}, \&
  {Zhang}}]{Zhang_BB-2007-Liang_EW-ApJ.666.1002Z}
{Zhang}, B.-B., {Liang}, E.-W., \& {Zhang}, B. 2007, \apj, 666, 1002

\bibitem[{{Zhang} \&
  {MacFadyen}(2009)}]{Zhang_W-2009-MacFadyen_A-ApJ.698.1261Z}
{Zhang}, W., \& {MacFadyen}, A. 2009, \apj, 698, 1261

\end{thebibliography}

\end{document}